\documentclass[11pt]{article}
\usepackage{amssymb}
\usepackage[symbol]{footmisc}
\usepackage[a4paper, total={6in, 8.3in}]{geometry}
\usepackage{sidecap}
\usepackage{multirow}
\usepackage[utf8]{inputenc}
\usepackage{latexsym}
\usepackage{caption}
\usepackage{subcaption}
\usepackage[hyperref]{xcolor}
\usepackage{epsfig}
\graphicspath{{projectplots/}}

\usepackage{tabularx} 
\usepackage{amsmath}  
\usepackage{graphicx} 
\usepackage{cite} 
\usepackage{hyperref} 
\hypersetup{colorlinks,breaklinks,
    colorlinks=true,
    linkcolor=blue,
    filecolor=magenta,      
    urlcolor=teal,
    citecolor=blue}
\def\beq{\begin{equation}}
\def\eeq{\end{equation}}
\def\bea{\begin{eqnarray}}
\def\eea{\end{eqnarray}}

\def\eq$#1${\begin{equation}#1\end{equation}}
\def\gat$#1${\begin{gather}#1\end{gather}}
\def\bal$#1${\begin{align}#1\end{align}}
\def\eqarr$#1${\begin{eqnarray}#1\end{eqnarray}}

\usepackage{relsize}
\usepackage{indentfirst}
\usepackage{amssymb}
\usepackage{bookmark}
\usepackage{amsthm}
\usepackage{amsmath}

\newcommand{\newc}{\newcommand}
\def\eq$#1${\begin{equation}#1\end{equation}}
\def\gat$#1${\begin{gather}#1\end{gather}}
\def\bal$#1${\begin{align}#1\end{align}}
\def\eqarr$#1${\begin{eqnarray}#1\end{eqnarray}}
\newc{\pa}{\partial}
\newc{\alp}{\alpha}
\newc{\gam}{\gamma}
\newc{\Gam}{\Gamma}
\newc{\del}{\delta}
\newc{\eps}{\epsilon}
\newc{\lam}{\lambda}
\newc{\sig}{\sigma}
\newc{\ups}{\upsilon}
\newc{\ome}{\omega}
\newc{\pphi}{\varphi}
\newc{\nonum}{\nonumber}
\newc{\hami}{\text{\textbf{\lat{H}}}}
\newc{\gren}{\mathcal{G}}
\newc{\lagr}{\mathcal{L}}
\newc{\timor}{\mathcal{T}}
\newc{\prop}{\mathcal{K}}
\newc{\zcal}{\mathcal{Z}}
\newc{\operx}{\text{\textbf{\lat{x}}}}
\newc{\opera}{\text{\textbf{\lat{a}}}}
\newc{\operp}{\text{\textbf{\lat{p}}}}
\newc{\operl}{\text{\textbf{\lat{L}}}}
\newc{\gfv}{g^{(5)}}
\newc{\kfv}{\kappa_{(5)}}
\newc{\tf}{\tilde{f}}
\newc{\tlam}{\tilde{\Lambda}}
\newc{\tl}{\tilde{\lam}}
\newc{\dist}{\displaystyle}
\newc{\ra}{\rightarrow}
\newc{\Ra}{\Rightarrow}

\begin{document}

\begin{titlepage}

\vspace*{1cm}
\begin{center}
{\bf \Large New Black-String Solutions for an Anti-de Sitter Brane
in\vspace{1mm} Scalar-Tensor Gravity}

\bigskip \bigskip \medskip

{\bf Theodoros Nakas},$^{1,}$\footnote[1]{Email: thnakas@cc.uoi.gr} 
{\bf Nikolaos Pappas},$^{2,}$\footnote[2]{Email: npappas@cc.uoi.gr} and
{\bf Panagiota Kanti}$^{1,}$\footnote[3]{Email: pkanti@cc.uoi.gr} 

\bigskip
$^1${\it Division of Theoretical Physics, Department of Physics,\\
University of Ioannina, Ioannina GR-45110, Greece}

\bigskip 
$^2${\it Nuclear and Particle Physics Section, Physics Department,\\
National and Kapodistrian University of Athens, Athens GR-15771, Greece}

\bigskip \medskip
{\bf Abstract}
\end{center}
We consider a five-dimensional theory with a scalar field non-minimally coupled
to gravity, and we look for novel black-string solutions in the bulk.
By appropriately choosing the non-minimal coupling function of the scalar field, 
we analytically solve the gravitational and scalar-field equations in the bulk
to produce black-string solutions that describe a Schwarzschild-anti-de Sitter
space-time on the brane. We produce two complete such solutions that are both
characterised by a regular scalar field, a localised-close-to-our brane
energy-momentum tensor and a negative-definite, non-trivial bulk potential that
may support by itself the warping of the space-time even in the absence of the
traditional, negative, bulk cosmological constant. Despite the infinitely long
string singularity in the bulk, the four-dimensional effective theory on the
brane is robust with the effective gravity scale being related to the fundamental
one and the warping scale. It is worth noting that if we set the mass of
the black hole on the brane equal to zero, the black string disappears leaving
behind a regular brane-world model with only a true singularity at the boundary of
the fifth dimension.

\end{titlepage}

\renewcommand*{\thefootnote}{\arabic{footnote}}


\section{Introduction}

The General Theory of Relativity \cite{einstein1, einstein2, einstein3} is a mathematically
beautiful, tensorial theory of gravity that predicts a variety of fascinating
gravitational objects such as black holes, wormholes or compact stars. It was
formulated in the (3+1) space-time dimensions of our world but it has since been
extended to an arbitrary number of extra dimensions \cite{Kaluza:1921tu, Klein:1926tv,
Misha, Akama, ADD1, ADD2, ADD3, RS1, RS2}. Although the mathematical extension
of the theory to a higher number of dimensions was straightforward, the implications
on the physical properties of the derived solutions have been quite dramatic. Black-hole
solutions are a characteristic example of such a transition: whereas in four dimensions
these solutions have been uniquely determined and classified (see, for example, 
\cite{MTW, Carter}), in higher number of dimensions, they are merely one species
of the family of the so-called black objects, which includes black holes, black strings,
black branes, black rings, or black saturns \cite{Emparan-review}.
The study of the properties of all these objects is still a topic of intensive research
activity and the formulation of uniqueness theorems or their complete classification
is still lacking. 

In fact, the determination of regular black-hole solutions in the context of the
so-called brane-world models \cite{RS1, RS2} has been a challenging task that, in
part, remains still unfulfilled. These models present the additional complication
of the presence of the self-energy of the brane, which is one of the fundamental
features of the theory as it gives rise to the desired warping of space-time.
This feature has so far prevented the determination of an analytical, non-approximate
solution describing a regular, localized-close-to-our-brane black hole. Although
a large number of different theoretical approaches and techniques have been used 
(see Refs. \cite{CHR, EHM1, EHM2, tidal, KT, Papanto, CasadioNew, Frolov, KOT,
Kofinas, Karasik, GGI, CGKM, Cuadros, AS, Heydari, Andrianov, Ovalle1, KPZ, Ovalle2,
Harko, Ovalle3, daRocha1, daRocha2, Ovalle4, KPP2, review1, review2, review3,
review4, review5, review6, review7, Nakas} for an impartial list of works),
only numerical solutions describing, in principle, regular black holes have been found
\cite{KTN, Kudoh, TT, Kleihaus, FW, Page1, Page2}. It is worth noting that,
in the context of theories where the brane self-energy may be ignored \cite{ADD1,
ADD2, ADD3}, analytical forms of higher-dimensional black holes were derived
long ago \cite{Tangherlini, MP} and the study of their properties, either
classical or quantum, has led to a separate direction of research activity. 

Whereas analytical solutions describing higher-dimensional black holes in a warped
space-time are still elusive, the other members of the family of black objects have
been easier to find \cite{Emparan-review}. For example, the first black-string solution
in the context of the models \cite{RS1, RS2} was derived in \cite{CHR}: although
from the point of view of a four-dimensional observer that solution described a
black hole on the brane, the complete bulk solution had a string-like singularity
extending all the way from the brane to the boundaries of the fifth dimension. 
This non-physical gravitational solution was, in addition, plagued by an intrinsic
instability that segmented the infinitely-extended black string to a sequence of
black cigars \cite{GL, RuthGL}. Since then, a variety of higher-dimensional 
black strings have been derived in the literature---see, for example Refs. 
\cite{Gubser, Wiseman, Kudoh2, Sorkin1, Sorkin2, Kleihaus2, Headrick,
Charmousis1, Charmousis2, Figueras, Kalisch, Emparan2, Cisterna1, KNP,
Cisterna2, Cisterna3}. 

The easiness with which classical black-string solutions arise in the context of
a brane-world model, as opposed to the difficulty that we face when we look for
black-hole solutions, has fuelled a great controversy in the literature over the
years \cite{Bruni, Dadhich, Tanaka, EFK, EGK, Fitzpatrick, Zegers, Yoshino, Dai},
which in fact remains undecided in the absence of an analytical solution describing
a regular black hole. Let us consider, for example, a brane-world model that
contains a bulk scalar field with an arbitrary potential and a non-minimal coupling
to gravity. Brane-world solutions with a Minkowski space-time on our
brane\footnote{Brane-world models based on a five-dimensional bulk scalar field
leading to a cosmologically evolving brane were also studied in \cite{Bhatta1, Bhatta2}.} were
studied in the context of this theory in \cite{Farakos1, Bogdanos1, Farakos2, Farakos3}.
In \cite{KPZ, KPP2}, the same theory was exhaustively studied for the determination
of an analytical solution describing a regular black hole localised close to our brane;
apart from the scalar degree of freedom that was supplemented by an arbitrary bulk
potential, a generalised ansatz for the higher-dimensional line-element was employed
that allowed for a diverse, and even dynamical, black-hole solution on the brane.
Yet, no viable complete bulk solution was found.

In contrast, the analyses of \cite{KPZ, KPP2} have shown that a scalar-tensor
theory of gravity in a five-dimensional warped space-time may allow quite easily
for new black-string solutions to emerge. Indeed, in \cite{KNP}, we focused on
such a theory and looked for analytical bulk solutions of the complete set of
gravitational and scalar-field equations. The solution for the five-dimensional
line-element described indeed a black string reducing to a Schwarzschild-(anti-)de
Sitter space-time on the brane. We chose to study the case of a positive cosmological
constant on the brane, which remarkably resulted in the condition that the non-minimal
coupling function of the scalar field to gravity had to be negative over a particular
regime in the bulk. Despite the anti-gravitating force that that condition led to,
some of the solutions found had a robust four-dimensional effective theory on the
brane. 

In this work, we focus on the case of a negative cosmological constant on the brane,
which as we will demonstrate removes the condition of the negative sign of the 
non-minimal coupling function in the bulk. With the gravity thus having everywhere
the correct sign, we will look again for analytical solutions describing novel black
strings. We will explicitly solve the coupled system of gravitational and scalar-field
equations to determine both the bulk gravitational background and the scalar field
configuration. Demanding the regularity of the scalar field everywhere in the bulk,
we will reduce the general form of the coupling function to two particular choices.
Both choices lead to analytical black-string solutions that, apart from the infinitely-long
string singularity, are free of any additional bulk singularities associated either with
the scalar field or with the five-dimensional line-element. The solutions that we
found exhibit
also a number of attractive features: the energy-momentum tensor of the theory is
everywhere regular and localised close to our brane leading to a five-dimensional
Minkowski space-time at large distances away from it. Also, the warping of the fifth
dimension may be supported exclusively by the negative-definite, non-trivial bulk
potential of the scalar field, a result which makes redundant the presence of the
negative bulk cosmological constant. Finally, the five-dimensional theory leads to
a robust four-dimensional effective theory on the brane with the effective gravity
scale being related to the fundamental one by a relation almost identical to the 
one appearing in \cite{RS1, RS2}. It is worth noting that if we set the mass of
the black hole on the brane equal to zero, the black string disappears leaving
behind a regular brane-world model with only a true singularity at the boundary of
the fifth dimension.

Our paper has the following outline: in Sec. 2, we present our theory, the field
equations and set a number of physical constraints on the scalar field and its
coupling function. In Sec. 3, we study in detail the case of an exponential coupling
function and determine the complete bulk solution, its physical properties, the 
junction conditions as well as the effective theory on the brane. We repeat the
analysis for another interesting case, that of a double exponential coupling
function, and discuss its properties in Sec. 4. We finally present our conclusions
in Sec. 5.


\section{The Theoretical Framework}

In this work, we consider a five-dimensional gravitational theory that
contains the five-dimensional scalar curvature $R$, a bulk cosmological
constant $\Lambda_5$, and a five-dimensional scalar field $\Phi$
with a self-interacting potential $V_B(\Phi)$. The scalar field also
possesses a non-minimal coupling to $R$ via a general coupling function
$f(\Phi)$: this function will be initially left arbitrary and will be 
given indicative specific forms when particular physical criteria are set
towards the end of this section. Thus, the action functional of the
{\it class} of theories we consider in this work is given by the expression
\beq
\label{action}
S_B=\int d^4x\int dy \,\sqrt{-g^{(5)}}\left[\frac{f(\Phi)}{2\kappa_5^2}R
-\Lambda_5-\frac{1}{2}\,\pa_L\Phi\,\pa^L\Phi-V_B(\Phi)\right].
\eeq
In the above, $g^{(5)}_{MN}$ is the metric tensor of the five-dimensional
space-time, and $\kappa_5^2=8\pi G_5$ incorporates the five-dimensional gravitational
constant $G_5$. Our four-dimensional world is represented by a 3-brane
located at $y=0$ along the fifth spatial dimension. Thus, in order to complete
the model, the following brane action should be added to the bulk one,
presented above,
\beq
\label{action_br}
S_{br}=\int d^4x\sqrt{-g^{(br)}}(\lagr_{br}-\sigma)=
-\int d^4x\int dy\sqrt{-g^{(br)}}\,[V_b(\Phi)+\sigma]\,\delta(y)\,.
\eeq
For simplicity, the quantity $\lagr_{br}$, which is related to the matter/field
content of the brane, contains only an interaction term $V_b(\Phi)$ of the bulk
scalar field with the brane. In addition, $\sigma$ is the constant brane self-energy,
and $g^{(br)}_{\mu\nu}=g_{\mu\nu}^{(5)}(x^\lam,y=0)$ is the induced-on-the-brane
metric tensor. Throughout this work, we will denote five-dimensional indices with
capital Latin letters $M,N,L,...$ and four-dimensional indices with lower-case
Greek letters $\mu,\nu,\lambda,...$ as usual.

If we vary the complete action $S=S_B+S_{br}$ with respect to the scalar field,
we obtain its equation of motion
\beq
-\frac{1}{\sqrt{-g^{(5)}}}\,\pa_M\left(\sqrt{-\gfv}g^{MN}\pa_N\Phi\right)=
\frac{\pa_\Phi f}{2 \kappa^2_5} R-\pa_\Phi V_B 
-\frac{\sqrt{-g^{(br)}}}{\sqrt{-\gfv}}\,\partial_\Phi V_b\,\delta(y)\,\,.
\label{phi-eq-0}
\eeq
The above must be supplemented by the gravitational field equations, which
follow from the variation of the action with respect to the metric-tensor
components $g^{(5)}_{MN}$, and have the form
\eq$\label{grav_eqs}
f(\Phi)\,G_{MN}\sqrt{-g^{(5)}}=\kappa_5^2\left[(T^{(\Phi)}_{MN}-g_{MN}\Lambda_5)
\sqrt{-g^{(5)}}-[V_b(\Phi)+\sigma]\,g^{(br)}_{\mu\nu} \delta^\mu_M\delta^\nu_N\delta(y)\sqrt{-g^{(br)}}\right],$
where
\eq$\label{Tmn}
T^{(\Phi)}_{MN}=\pa_M\Phi\,\pa_N\Phi+g_{MN}\left[-\frac{\pa_L\Phi\pa^L\Phi}{2}-V_B(\Phi)\right]
+\frac{1}{\kappa_5^2}\left[\nabla_M\nabla_Nf(\Phi)-g_{MN}\Box f(\Phi)\right].$

In order to derive the explicit form of the aforementioned field equations 
(\ref{phi-eq-0}) and (\ref{grav_eqs}), we need to specify the form of the
five-dimensional gravitational background. We assume that this has the
form of a warped space-time with its four-dimensional part having the form
of a generalised Vaidya line-element. Thus, we consider the expression
\eq$\label{metric}
ds^2=e^{2A(y)}\left\{-\left[1-\frac{2m(r)}{r}\right]dv^2+2dvdr+r^2(d\theta^2+
\sin^2\theta d\varphi^2)\right\}+dy^2\,.$
In the above, the function $e^{2A(y)}$ stands for the warp factor along the fifth
dimension, and $m(r)$ is a generalised mass function. A first advantage of the above
metric form is that, upon setting the coordinate $y$ along the extra dimension to zero,
the part inside the curly brackets reduces to the line-element on the brane. Then, 
if the function $m(r)$ is assumed to be a constant $M$, the four-dimensional
line-element is merely the Vaidya transformation of the Schwarzschild solution,
where $M$ is the black hole mass. Despite the fact that the line-element on the
brane admits such an attractive interpretation, the complete five-dimensional
line-element, in the context of the purely gravitational Randall-Sundrum model
\cite{RS1, RS2}, was shown to describe a black-string \cite{CHR} plagued also by
instabilities \cite{GL, RuthGL}. 

Since then, and despite all efforts, no analytical, closed form of a regular,
localised-on-the-brane black-hole solution has been found. During that quest,
line-element (\ref{metric}) was used in a number of works \cite{KOT, KPZ,
KPP2} as it exhibited a second advantage: since it contained no horizon in 
its four-dimensional part, it did not lead to additional bulk singularities
\cite{KT, KOT}. A generalized Vaidya form, where $m$ is not a constant 
any more but a function of the coordinates, was employed in an effort to
increase the flexibility of the model and allow for brane black-hole solutions
to deviate from the overly simple Schwarzschild one. In fact, the line-element
(\ref{metric}) with $m=m(v,r,y)$ was used in \cite{KPP2} in the context of the
scalar-tensor theory (\ref{action}) but, despite its flexibility, which was further
increased by the scalar degree of freedom, led to no robust black-hole solutions. 

As in our previous work \cite{KNP}, here, we also turn our attention to the 
derivation of novel black-string solutions that the theory (\ref{action}) 
seems to possess in abundance. Indeed, in \cite{KNP} we demonstrated that
the theory allows for a variety of such solutions that may be constructed
analytically. Assuming that the cosmological constant on the brane is non-vanishing
and positive, novel black-string solutions were derived and characterised
by interesting properties such as the existence of an anti-gravitating
regime in the bulk. Here, we continue our pursuit for novel black strings
in the context of the theory (\ref{action}) but allow the brane cosmological
constant to be negative. We will again make the choice $m=m(r)$ for the
mass function, and demonstrate by direct integration that a class of 
black-string solutions arises that, on the brane, has the interpretation
of a Schwarzschild-anti-de Sitter black hole. 

Therefore, employing line-element (\ref{metric}) and the relation
$\sqrt{-g^{(5)}}=\sqrt{-g^{(4)}}$ that holds in this case, the scalar-field
equation of motion (\ref{phi-eq-0}) in the bulk [i.e. we ignore for now
the brane $\delta(y)$-term] takes the explicit form
\beq \label{phi-eq}
\Phi'' + 4A' \Phi' =\pa_\Phi f \left(10A'^2+4A''-e^{-2A}\frac{2\pa_rm+
r\,\pa_r^2m}{r^2}\right) +\pa_\Phi V_B\,.
\eeq
In the above, a prime ($'$) denotes the derivative with respect to the $y$-coordinate.
We have also assumed that the bulk scalar field depends only on the coordinate
along the fifth dimension, i.e. $\Phi=\Phi(y)$. For the explicit form of
the gravitational equations, we need the non-vanishing components of the Einstein
$G^{M}{}_N$ and energy-momentum $T^{(\Phi)M}{}_N$ tensors. In mixed form,
these are:
\bea
&~&G^0{}_0=G^1{}_1=6A'^2+3A''-\frac{2e^{-2A}\pa_rm}{r^2},\nonumber \\[1mm] 
&~&G^2{}_2=G^3{}_3=6A'^2+3A''-\frac{e^{-2A}\pa_r^2m}{r}, \label{GMN} \\[2mm] 
&~&G^4{}_4=6A'^2-\frac{e^{-2A}\left(2\pa_rm+r\pa_r^2m\right)}{r^2}, \nonumber
\eea
and  
\begin{gather}
T^{(\Phi)0}{}_0=T^{(\Phi)1}{}_1=T^{(\Phi)2}{}_2=T^{(\Phi)3}{}_3=
A' \Phi'\,\pa_\Phi f+\lagr_\Phi-\Box f, \nonumber \\[2mm]
T^{(\Phi)4}{}_4=(1+\pa_\Phi^2 f)\Phi'^2+\Phi''\,\pa_\Phi f+\lagr_\Phi-\Box f,
\label{TMN-mixed}
\end{gather}
respectively. Above, we have defined the quantities
\beq\label{Lagr}\lagr_{\Phi}=-\frac{1}{2}\,\pa_L\Phi\,\pa^L\Phi-V_B(\Phi)
=-\frac{1}{2}\,\Phi'^2-V_B(\Phi).
\eeq
and
\beq
\label{Box_f}
\Box f=4A' \Phi'\,\pa_\Phi f+\Phi'^2\,\pa_\Phi^2 f+\Phi''\,\pa_\Phi f.
\eeq
By substituting the aforementioned components of the Einstein and scalar energy-momentum
tensors in Eq. (\ref{grav_eqs}), and ignoring again the brane boundary terms, the
gravitational field equations in the bulk may be derived. We thus obtain three equations
that, upon some simple manipulation \cite{KNP}, assume the following form
\beq
\label{eq-mass}
r\,\pa_r^2m-2\pa_rm=0\,,
\eeq
\beq \label{grav-1}
f\left(3A''+e^{-2A}\frac{\pa_r^2m}{r}\right)=\pa_\Phi f \left(A'\Phi'-\Phi''\right)
-(1+\pa_\Phi^2 f)\Phi'^2\,,
\eeq
\beq
\label{grav-2}
f\left(6A'^2+3A''-\frac{2e^{-2A}\pa_rm}{r^2}\right)=A'\Phi'\,\pa_\Phi f+
\lagr_\Phi -\Box f-\Lambda_5\,.
\eeq
Note that, for notational simplicity, we have absorbed the gravitational constant
$\kappa_5^2$ in the expression of the general coupling function $f(\Phi)$.

As was explicitly shown in the Appendix B of \cite{KNP}, not all of Eqs.
(\ref{phi-eq}) and (\ref{eq-mass})-(\ref{grav-2}) are independent. Therefore, in
what follows, we will ignore Eq. (\ref{phi-eq}) and work only with the gravitational
equations. Of them, Eq. (\ref{grav-2}) serves to determine the scalar potential
in the bulk $V_B(\Phi)$. It is Eq. (\ref{grav-1}) that will provide the solution
for the scalar field $\Phi$ once the warp function $A(y)$, the mass function
$m(r)$, and the non-minimal coupling function $f(\Phi)$ are determined. For the
warp factor, we will make the assumption that this is given by the well-known
form $A(y)=-k |y|$ \cite{RS1, RS2}, with $k$ being a positive constant, as this ensures
the localization of gravity near the brane. The form of the mass function $m(r)$
readily follows by direct integration of Eq. (\ref{eq-mass}) that leads to the
expression
\beq
m(r)=M+ \Lambda r^3/6\,, \label{mass-sol}
\eeq
where $M$ and $\Lambda$ are arbitrary integration constants (the numerical
coefficient 1/6 has been introduced for later convenience). Substituting the
above form into line-element (\ref{metric}) and setting $y=0$, we may
easily see that the projected-on-the-brane background is given by the
expression
\eq$\label{metric-brane}
ds^2_4=-\left(1-\frac{2M}{r}-\frac{\Lambda r^2}{3}\right)dv^2+2dv dr+
r^2(d\theta^2+\sin^2\theta\ d\varphi^2)\,.$
In \cite{KNP}, we explicitly demonstrated that the above Vaidya form of the 
four-dimensional line-element may be transformed to the usual 
Schwarzschild-(anti-)de Sitter solution by an appropriate coordinate
transformation. Therefore, the arbitrary parameter $M$ is the mass of
the black-hole that the four-dimensional observer sees and $\Lambda$
the cosmological constant on the brane. 

The case of a positive cosmological constant on the brane (i.e. $\Lambda>0$)
was studied in our previous work \cite{KNP}; in the context of the present
analysis, we will focus on the case of a negative four-dimensional cosmological
constant ($\Lambda < 0$). Employing the form of the mass function (\ref{mass-sol})
and the exponentially decreasing warp factor\footnote{We assume a
${\bf Z}_2$-symmetry in the bulk under the change $y \rightarrow -y$; therefore,
henceforth, we focus on the positive $y$-regime.} $e^{2A(y)}=e^{-2ky}$,
Eq. (\ref{grav-1}) takes the form
\eq$\label{anti-grav-1}
(\Phi')^2=-\pa_y^2 f-k\pa_yf - \Lambda e^{2ky}f\,.$
In the above, we have also used the relations
\eq$\label{dif-f}
\pa_yf=\Phi' \,\pa_\Phi f, \hspace{1.5em}\pa_y^2f=
\Phi'^2\,\pa_\Phi^2 f+\Phi''\,\pa_\Phi f\,.$
The l.h.s of Eq. (\ref{anti-grav-1}) is positive; therefore the same should hold
for the r.h.s, too. Note that, for $\Lambda>0$, Eq. (\ref{anti-grav-1}) demands
that, at least at $y \rightarrow \infty$, the coupling function $f(\Phi)$ should
be negative for the scalar field to have a real first derivative there. Indeed,
in \cite{KNP}, we presented two analytic solutions of this theory where $f<0$
either far away from our brane or in the entire bulk regime. In contrast, in
the present case, where $\Lambda<0$, no such behaviour is necessary; thus in order
to have a normal gravity over the entire five-dimensional space-time, we
assume that $f(\Phi)$ is positive everywhere. 

In order to have a physically acceptable behaviour, a few more properties should
be assigned to the functions $\Phi=\Phi(y)$ and $f=f[\Phi(y)]$. Both functions
should, of course, be real and finite in their whole domain and of class $C^{\infty}$.
At $y\ra \infty$, both functions should satisfy the following relations, otherwise
the finiteness of the theory at infinity cannot be ascertained,
\eq$\label{con.1}
\lim_{y\ra\infty}\frac{d^n[f(y)]}{dy^n}=0,\hspace{1.5em}\forall n\geq 1,$
\eq$\label{con.2}
\lim_{y\ra\infty}\frac{d^n[\Phi(y)]}{dy^n}=0,\hspace{1.5em}\forall n\geq 1.$
These constraints guarantee that all components of the energy-momentum tensor
$T^{(\Phi)M}{}_N$ will be real and finite everywhere, and, in addition, localised
close to our brane. Then, demanding also the finiteness and the vanishing of the
r.h.s. of Eq. (\ref{anti-grav-1}) due to the constraint (\ref{con.2}), we conclude
that the coupling function $f(y)$ should, at infinity, decrease
faster~\footnote{Note that allowing the coupling function to vary exactly as
$e^{-2ky}$, i.e. $f(y)=f_0\,e^{-2ky}$, would lead to a finite, constant
value of $\Phi'^2$ at infinity, namely, $\Phi'^2_\infty=-f_0\,\Lambda>0$. This would
amount to having a diverging field at the boundary of spacetime but nevertheless
finite, constant values for the components of the energy-momentum tensor. We will
come back to this point later.}
than $e^{-2ky}$, i.e. $f(y)$ should be of the form:
\eq$\label{anti-f}
f(y)=f_0\ e^{g(y)},\hspace{1.5em} \left\{\begin{array}{c}
f_0>0\\ \\
g(y\ra\infty)<-2ky 
\end{array}\right\}.$
Consequently, upon integrating Eq. (\ref{anti-grav-1}), the following expression
is obtained for the scalar field:
\eq$\label{anti-phi}
\Phi(y)=\pm \sqrt{f_0}\int dy\ e^{\frac{g(y)}{2}}\sqrt{\tlam^2 e^{2ky}-g''-g'^2-kg'},$
where, for convenience, we have also set $\Lambda=-\tlam^2$. In order to proceed
further, we need to determine the exact form of the function $g(y)$. As we are
interested in deriving analytical solutions for both functions $f(y)$ and $\Phi(y)$,
the function $g(y)$ should have a specific form in order to result in a solvable
integral on the r.h.s. of Eq. \eqref{anti-phi}. Therefore, we make the
following two choices:
\gat$\label{anti-g1}
g(y)=-\lam ky,\hspace{1.5em}\lam\in(2,+\infty)\,,\\ \nonum\\
\label{anti-g2}
g(y)=-\mu^2 e^{\lam y},\hspace{1.5em}\left\{\begin{array}{c}\lam\in(0,+\infty)\\
\\ \mu\in\mathbb{R}\setminus \{0\}\end{array}\right\}.
$
The aforementioned expressions for $g(y)$ ensure that both $f(y)$ and $\Phi(y)$
have the desired properties outlined above and, in addition, lead to analytical
solutions. In the following sections, these two different cases will be studied
separately.


\section{The Simple Exponential Case}
\label{anti1}

We will start with the simple exponential case (\ref{anti-g1}), and derive first
the form of the scalar field and its potential in the bulk. We will then study their
main characteristics in terms of the free parameters of the model, and finally
address the effect of the junction conditions and the form of the effective 
theory on the brane. 

\subsection{The bulk solution}

\par In this case, we have $f(y)=f_0\ e^{-\lam k y}$, with $f_0>0$ and $\lam>2$.
Then, from Eq. (\ref{anti-grav-1}), we obtain
\beq
\Phi'^{\,2}(y)=f(y)\,(\tlam^2 e^{2ky}-\lam^2k^2+\lam k^2) \geq 0\,.
\label{anti1-Phi/f}
\eeq
For a non-zero and positive $f(y)$, the above inequality demands that the combination
inside the brackets should be positive. As this is an increasing function of $y$, it
suffices to demand that this holds at the location of the brane, at $y=0$. 
Then, we obtain the following constraint on the parameters of the theory:
\beq
\label{anti1-con1}
\frac{\tlam^2}{\lam(\lam-1)k^2}> 1\,.
\eeq
The function $\Phi'^{\,2}(y)$ could, in principle, be zero at the point where
$\Phi(y)$ has an extremum. However, from Eq. \eqref{anti1-Phi/f}, we may easily
see that this may happen only at $y_0=\frac{1}{2k}\ln\left(\frac{\lam(\lam-1)k^2}
{\tlam^2}\right)$, which, upon using Eq. (\ref{anti1-con1}), turns out to be
negative. Therefore, the scalar field does not have any extremum in the whole
domain $0 \leq y < \infty$, which in turn means that $\Phi(y)$ is a one-to-one
function in the same region. The $\mathbf{Z}_2$ symmetry of the extra dimension
ensures that this result holds in the region $y<0$ as well. We note this
property for later use.

\par Equation \eqref{anti-grav-1} can be rewritten as
\bea(\Phi')^2 &=&f_0\lam(\lam-1)k^2\left[\frac{\tlam^2}{\lam(\lam-1)k^2}e^{2ky}
-1\right]e^{-\lam ky} \nonum\\[3mm]
&=& f_0\lam(\lam-1)k^2\left[\frac{\tlam^2}{\lam(\lam-1)k^2}\right]^{\lam/2}(w-1)\,w^{-\lam/2}\,,
\label{phi'-w}
\eea
where we have introduced the new variable $w$ via the definition
\eq$\label{anti1-w}
w(y)\equiv \frac{\tlam^2 e^{2 k y}}{\lam(\lam-1)k^2}\,.$
Because of the constraint \eqref{anti1-con1}, it is obvious that $w(y)$ is greater than
unity for all values of the extra coordinate $y$. Then, applying the chain rule to
the l.h.s. of Eq. (\ref{phi'-w}) and integrating, we obtain for the scalar field
the integral expression 
\gat$
\Phi(w)=\pm\frac{\sqrt{f_0\lam(\lam-1)}}{2}\left[\frac{\tlam^2}{\lam(\lam-1)k^2}
\right]^{\lam/4}\int dw\ (w-1)^{\frac{1}{2}}\,w^{-\frac{\lam}{4}-1}\,.$

In order to evaluate the above integral, we perform a second change of variable;
namely, we set $w=1/(1-z)$. Then,
\gat$\int dw\ (w-1)^{\frac{1}{2}}w^{-\frac{\lam}{4}-1}=
\int dz\ z^{\frac{1}{2}}(1-z)^{\frac{\lam}{4}-\frac{3}{2}}=\int_0^z dt\,
t^{\frac{1}{2}}(1-t)^{\frac{\lam}{4}-\frac{3}{2}}+C_1\,,$
where an arbitrary constant $C_1$ has been introduced in order to set the lower
boundary value of the integral equal to zero. Finally, by employing the rescaled variable 
$t'=t/z$, the above integral takes its final form
\gat$
\label{integral-zt}
z^{\frac{3}{2}}\int_0^1 dt'\ t'^{\frac{1}{2}}(1-zt')^{\frac{\lam}{4}-\frac{3}{2}}+C_1=
\frac{2}{3}\left(\frac{w-1}{w}\right)^{3/2}\,_2F_1\left(\frac{3}{2}-
\frac{\lam}{4},\frac{3}{2};\frac{5}{2};\frac{w-1}{w}\right)+C_1\,,$
where we used the integral representation of the hypergeometric function \cite{Abramowitz}
\beq
\,_2F_1\left(a,b;c;z\right)=\frac{\Gamma(c)}{\Gamma(b)\Gamma(c-b)}\int_0^1 dt'\
t'^{b-1}(1-t')^{c-b-1}(1-zt')^{-a},\hspace{1.5em}Re(c)>Re(b)>0\,.
\eeq
We now observe that $\Phi$ appears in the field equations \eqref{phi-eq}, \eqref{grav-1}
and \eqref{grav-2} only through the coupling function $f(\Phi)$ and the bulk potential
$V_B(\Phi)$. Therefore, any shift in the value of the scalar field by an arbitrary
constant would result in a change in the value of $f$ by a constant amount that could
nevertheless be reabsorbed in the redefinition of the value of the arbitrary coefficient
$f_0$; the value of the bulk potential $V_B$ would also change by a constant amount but this
could again be reabsorbed in the value of the arbitrary bulk cosmological constant
$\Lambda_5$. Due to this translation symmetry with respect to the value of the scalar
field $\Phi(y)$, we may set the arbitrary constant $C_1$ in Eq. (\ref{integral-zt})
equal to zero. This brings the solution for the scalar field into its final form
\eq$\label{anti1-phi}
\Phi_{\pm}(y)=\pm\frac{\sqrt{f_0\lam(\lam-1)}}{3}\left[\frac{\tlam^2}{\lam(\lam-1)k^2}
\right]^{\lam/4}\left[\frac{w(y)-1}{w(y)}\right]^{3/2}\,_2F_1  \left(\frac{3}{2}-\frac{
\lam}{4},\frac{3}{2};\frac{5}{2};\frac{w(y)-1}{w(y)}\right).$ 

Although the function $w(y)$ is greater than unity for all values of the extra dimension $y$,
the argument $z=\frac{w-1}{w}$ of the hypergeometric function in the previous relation is
always positive and smaller than unity. Hence, we may use the well-known expansion for the 
hypergeometric function in power series,
\beq
\,_2F_1(a,b;c;z)=\,_2F_1(b,a;c;z)=\sum_{n=0}^{\infty}
\frac{a^{(n)}b^{(n)}}{c^{(n)}}\frac{z^n}{n!}\,,
\label{Expansion-F}
\eeq
where $|z|<1$, and the quantities of the form $q^{(n)}$ denote (rising) Pochhammer symbols,
namely,
\beq
q^{(n)}=\frac{\Gamma(q+n)}{\Gamma(q)}=
\left\{\begin{array}{cc}q(q+1)\cdots(q+n-1)\,, & n>0\\ \\ 1\,, & n=0\end{array}\right..
\eeq
Thus, we find
\gat$
\label{final-F}
\,_2F_1  \left(\frac{3}{2}-\frac{\lam}{4},\frac{3}{2};\frac{5}{2};\frac{w-1}{w}\right)=
\sum_{n=0}^\infty \frac{\Gamma\left(\frac{3}{2}-\frac{\lam}{4}+n\right)}{\Gamma\left(\frac{3}{2}-\frac{\lam}{4}\right)}\,
\frac{3}{(2n+3)n!}\left(\frac{w-1}{w}\right)^n,$
where we have also used the property $\Gamma(1+z)=z\Gamma(z)$.
There are two interesting categories of values for the parameter $\lam$ that lead to simple
and elegant expressions for the hypergeometric function and subsequently for the scalar field.
These are $\lam=2(1+2q)$ and $\lam=4q$, where $q$ is any positive integer.
Let us examine each case separately. 

\begin{itemize}
\item If $\lam=2(1+2q)$ with $q \in {\mathbb{Z}}^{>}$, then, from Eq. (\ref{final-F}),
we have:
\bal$
\,_2F_1  \left(\frac{3}{2}-\frac{\lam}{4},\frac{3}{2};\frac{5}{2};\frac{w-1}{w}\right)
&=\sum_{n=0}^{\infty}(1-q)^{(n)}\frac{3}{(2n+3)n!}\left(\frac{w-1}{w}\right)^n\nonum\\[2mm]
&=\left\{\begin{array}{ccr}
1\,, & & q=1\\[4mm]
1+\sum_{n=1}^{q-1}\frac{3(-q+1)(-q+2)\cdots(-q+n)}{(2n+3)n!}\left(\frac{w-1}{w}\right)^n, & & q>1
\end{array}\right\} .
\label{F-series2}$
In the second line of the above expression, the upper limit of the sum has been changed
from $\infty$ to $q-1$ since, for $q$ and $n$ positive, the sum will be trivial for any
value of $n$ equal or higher than $q$ due to the factor $(-q+n)$. As
indicative cases, we present below the form of the scalar field
for\footnote{For completeness, we present here also the solution for the limiting case
with $q=0$ (i.e. for $\lam=2$); this has the form
$$\Phi_{\pm}(y)=\pm\sqrt{f_0}\,\sqrt{\frac{\tlam^2}{k^2}}
\left[\,{\rm arctanh}\left(\sqrt{\frac{w-1}{w}}\right)-\sqrt{\frac{w-1}{w}}\,\right].$$
Although the field diverges at infinity---see footnote 3---the components of the
energy-momentum tensor exhibit a regular behaviour as we will comment later.}
$q=1$ (i.e. $\lambda=6$), 
\beq
\Phi_{\pm}(y)=\pm\frac{\sqrt{f_0}}{90}\left(\frac{\tlam^2}{k^2}\right)^{3/2}\left(\frac{w-1}{w}\right)^{3/2},
\eeq
and $q=2$ (i.e. $\lambda=10$),
\beq
\Phi_{\pm}(y)=\pm\frac{\sqrt{f_0}}{3\times 90^2}\left(\frac{\tlam^2}{k^2}\right)^{5/2}
\left(\frac{w-1}{w}\right)^{3/2}\left[1-\frac{3}{5}\left(\frac{w-1}{w}\right)\right].
\eeq
The above expressions follow easily by using Eqs. (\ref{anti1-phi}) and (\ref{F-series2})
and substituting the aforementioned values of the parameter $q$.

\item If $\lam=4q$ with $q \in {\mathbb{Z}}^{>}$, we can always express the hypergeometric function 
in Eq. (\ref{anti1-phi}) in terms of elementary functions, namely, 
$\arcsin\left(\sqrt{\frac{w-1}{w}}\right)$, square roots, and powers of the argument
$\frac{w-1}{w}$. The process that one follows to obtain this expression is
presented in detail in Appendix \ref{hyper-analysis}. Thus, for $\lam=4$ (i.e. $q=1$),
using Eq. (\ref{B.3}), we may straightforwardly write
\beq
\Phi_{\pm}(y)=\pm\frac{\sqrt{f_0}}{4\sqrt{3}}\ \frac{\tlam^2}{k^2}
\left(\frac{w-1}{w}\right)^{1/2}\left[\sqrt{\frac{w}{w-1}}
\arcsin\left(\sqrt{\frac{w-1}{w}}\right)-\sqrt{\frac{1}{w}}\ \right].
\label{Phi-lam4}
\eeq
For larger values of $\lambda$ (i.e. for $q=1+\ell$, with $\ell \in {\mathbb{Z}}^{>}$),
we  should use instead Eq. (\ref{hyper-lam-4q-final}) together with the
constraint (\ref{alpha-beta}). The latter, as outlined in Appendix \ref{hyper-analysis},
reduces to a set of linear equations that determine the unknown coefficients
$\alpha, \beta_1, \cdots, \beta_\ell$. For example, for $\ell=1$, the set of
equations that follow from Eq. (\ref{alpha-beta}) is
\beq
\{2\alpha-\beta_1=0\,, \qquad 2\alpha+3\beta_1=3\}\,,
\eeq
leading to the values $\alpha=3/8$ and $\beta=3/4$. Then, after substituting these in
Eq. (\ref{hyper-lam-4q-final}), the solution for $q=2$, or equivalently for
$\lam=8$, follows from Eq. (\ref{anti1-phi}) and has the form
\gat$\Phi_{\pm}(y)=\pm\frac{\sqrt{f_0}}{896\sqrt{14}}\ \frac{\tlam^4}{k^4}\left(\frac{w-1}{w}\right)^{1/2}\left[\sqrt{\frac{w}{w-1}}\arcsin\left(\sqrt{\frac{w-1}{w}}\right)+\sqrt{\frac{1}{w}}\left(1-\frac{2}{w}\right)\right].$
Solutions for larger values of $q$, and thus of $\lam$, may be derived in the
same way in terms again of analytic, elementary functions.
\end{itemize}
In all the above, particular expressions for the scalar field $\Phi_{\pm}$, which follow for
specific values of the parameter $\lam$, the dependence on the extra coordinate $y$ is easily
made explicit by employing Eq. (\ref{anti1-w}). Also, for all other values of 
$\lam \in {\mathbb{R}}^{>2}$, which do not fall in the aforementioned categories, the
scalar field may still be expressed in terms of the hypergeometric function through
Eqs. (\ref{anti1-phi}) and (\ref{final-F}).

\begin{figure}[t]
    \centering
    \begin{subfigure}[b]{0.48\textwidth}
        \includegraphics[width=\textwidth]{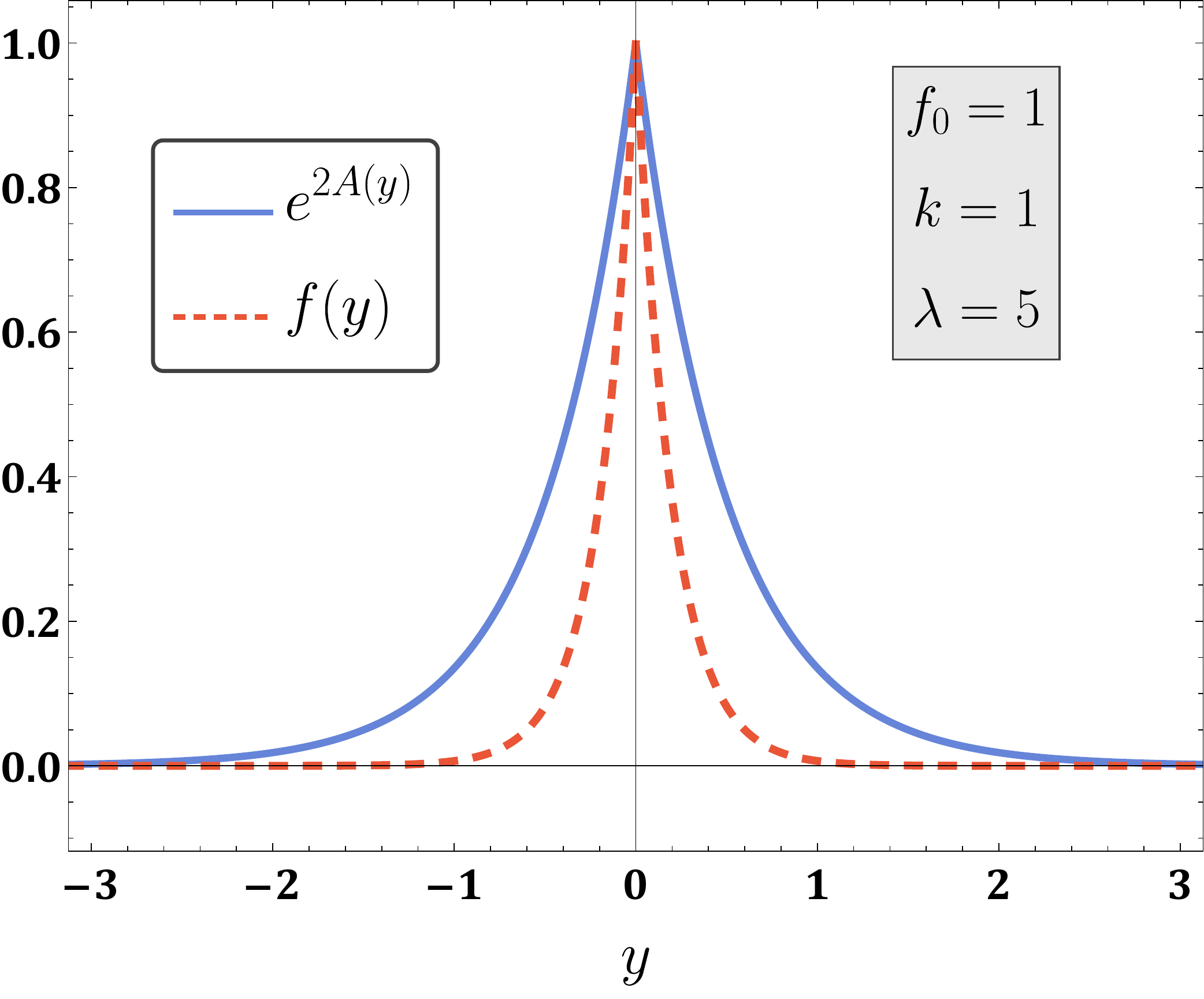}
        \caption*{\hspace{1.3em}(a)}
        \label{fig1a}
    \end{subfigure}
    ~ 
    \begin{subfigure}[b]{0.495\textwidth}
        \includegraphics[width=\textwidth]{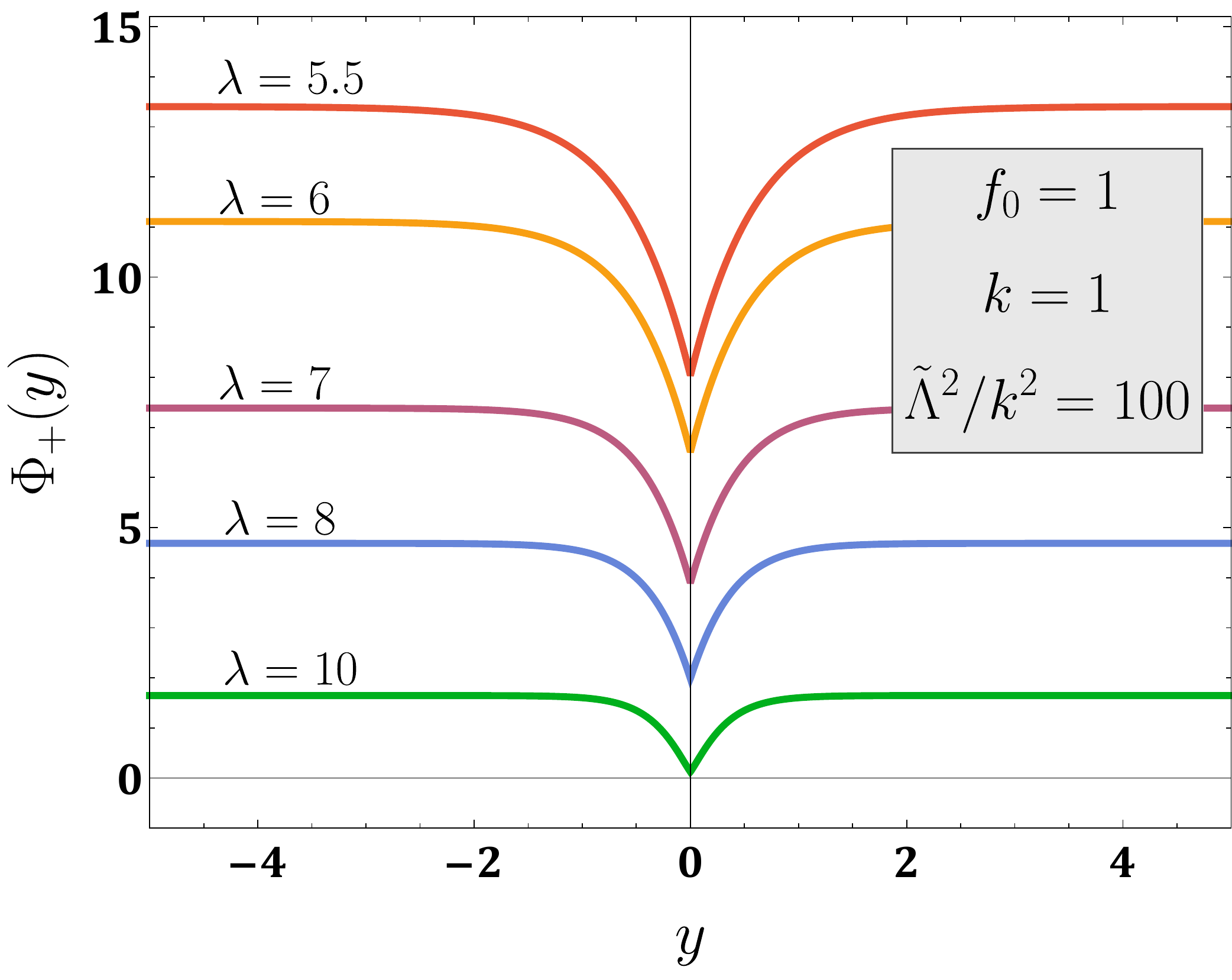}
        \caption*{\hspace{3em}(b)}
        \label{fig1b}
    \end{subfigure}
    ~ 
    \caption{(a) The warp factor $e^{2A(y)}=e^{-2k|y|}$ and the coupling function
    $f(y)=f_0\,e^{-\lam k |y|}$ in terms
    of the coordinate $y$, for $f_0=1$, $k=1$, $\lam=5$. (b) The scalar field $\Phi_{+}(y)$ also in
    terms of the coordinate $y$, for $f_0=1$, $k=1$, $\tlam^2/k^2=100$ and $\lam=5.5,6,7,8,10$ (from top to bottom).}
   \label{anti1_warp_f_Phi}
\end{figure}

\par Let us now investigate the physical characteristics of the solutions we have derived.
In Fig. 1(a), we depict the form of the warp factor $e^{-2k|y|}$ and the coupling function
$f(y)=f_0\,e^{-\lam k |y|}$ in terms of the coordinate $y$ along the fifth dimension,
for $f_0=1$, $k=1$ and $\lam=5$. The warp factor exhibits the anticipated localization
close to the brane while the non-minimal coupling function mimics this behaviour by
decreasing exponentially fast away from the brane and reducing to zero at the boundary of
spacetime. In fact, the larger the parameter $\lam$, the faster the decrease rate
of $f$ is; thus by increasing $\lam$, the non-minimal coupling of the scalar field 
to gravity is effectively ``localised'' closer to the brane. 

Figure 1(b) depicts
the scalar field $\Phi_+(y)$ for different values of the parameter $\lam$ and for $f_0=1$,
$k=1$, $\tlam^2/k^2=100$. It is straightforward to deduce from Fig. 1(b) that the scalar
field $\Phi_+(y)$ exhibits a reverse behaviour, compared to $f(y)$, by increasing away
from the brane and adopting a constant, non-vanishing value at the boundary of spacetime.
Note that, as $\lam$ increases, the scalar field reaches this constant asymptotic value
faster; that is, a more ``localised'' coupling function keeps also the non-trivial profile
of the scalar field closer to the brane. 
Overall, for $\lam > 2$, the scalar field presents a well-defined profile over the entire
extra dimension in accordance with the desired properties set in the previous section.
In Fig. 1(b), we chose to plot $\Phi_+(y)$, i.e. we chose the positive sign in
Eq. (\ref{anti1-phi}) for the expression of the scalar field. A second class of solutions
exists for $\Phi=\Phi_-$, with the only difference being Fig. 1(b) becoming its mirror
image with respect to the horizontal axis. The sign of the scalar field, however, does
not affect either the potential $V_B(y)$ or the components of the energy-momentum tensor,
as we will soon see. Finally, let us emphasize the fact that, as Fig. 1(b) reveals, the 
qualitative behaviour of the scalar field in terms of the parameter $\lam$ remains
unchanged. This holds despite the fact that the value of $\lam$ does affect the exact,
analytic expression of the scalar field, as we have shown in detail above; we may thus
conclude that solutions emerging for non-minimal coupling functions of a simple exponential
form, differing only in the value of the parameter $\lam$, i.e. in the decrease rate of
$f$ with $y$, lead to a class of black-string solutions with the same qualitative characteristics.

\par The potential of the field $V_B$ in the bulk can be determined from Eq. \eqref{grav-2}:
substituting the functions $m(r)$ and $A(y)$, we obtain
\beq
V_B=-\Lambda_5-\frac{1}{2}\,\Phi'^2 +3k \partial_y f -\partial_y^2 f -
f\left(6k^2 -\Lambda e^{2ky}\right),
\label{V_B-1}
\eeq
where we have also used the relations \eqref{dif-f}. Note that the bulk potential is indeed
insensitive to the sign of $\Phi_{\pm}$ that enters the above expression through $\Phi'^2$.
If we also employ Eq. \eqref{anti-grav-1} to substitute $\Phi'^2$, and use the exponential
form for $f(y)$, the following expression readily follows for the potential $V_B$ in terms
of the extra dimension $y$:
\eq$\label{anti1-pot}
V_B(y)=-\Lambda_5-\frac{k^2f_0}{2}\ e^{-\lam ky}\left(12+7\lam+\lam^2+\frac{3\tlam^2}{k^2}e^{2ky}\right).$
We observe that the combination $V_B(y)+\Lambda_5$, which appears in the action (\ref{action})
as well as in the components of the total energy-momentum tensor as we will shortly see, is
always negative definite. This combination, even for $\Lambda_5=0$, may therefore provide by
itself the negative distribution of energy in the bulk that is
necessary for the support of the AdS spacetime and the localisation of gravity. A similar result
was derived in \cite{KNP} where the case of a positive cosmological constant $\Lambda$ on
the brane was considered: there, the positive $\Lambda$ added a positive contribution to the
value of $V_B$, that was thus decreased in absolute value, while here the negative $\Lambda$
gives an extra boost to the negative value of $V_B$. The profile of the bulk potential of the
scalar field is depicted in Fig. 2(a) for $f_0=1/5$, $k=1$, $\tlam^2/k^2=9$ and $\lam=2.5$. 
The potential is everywhere finite and remains localized close to the brane. For all values of
$\lam>2$, it goes to zero with an exponential decay rate that increases with $\lam$. The
vanishing of $V_B$ at the boundary of spacetime, together with the similar behaviour of
the coupling function $f$ in the same regime and the constant value that the scalar field
assumes there, points to the conclusion that the non-minimally coupled scalar field, after
serving its purpose of localising gravity close to the brane, completely disappears leaving
behind a 5-dimensional Minkowski spacetime. Only, for $\lam=2$, an asymptotic bulk
cosmological constant equal to $-3f_0 \tilde \Lambda^2/2$ remains, thus leading to an
asymptotically AdS spacetime, but only by paying the price of a diverging scalar field at
infinity.

\begin{figure}[t!]
    \centering
    \begin{subfigure}[b]{0.45\textwidth}
        \includegraphics[width=\textwidth]{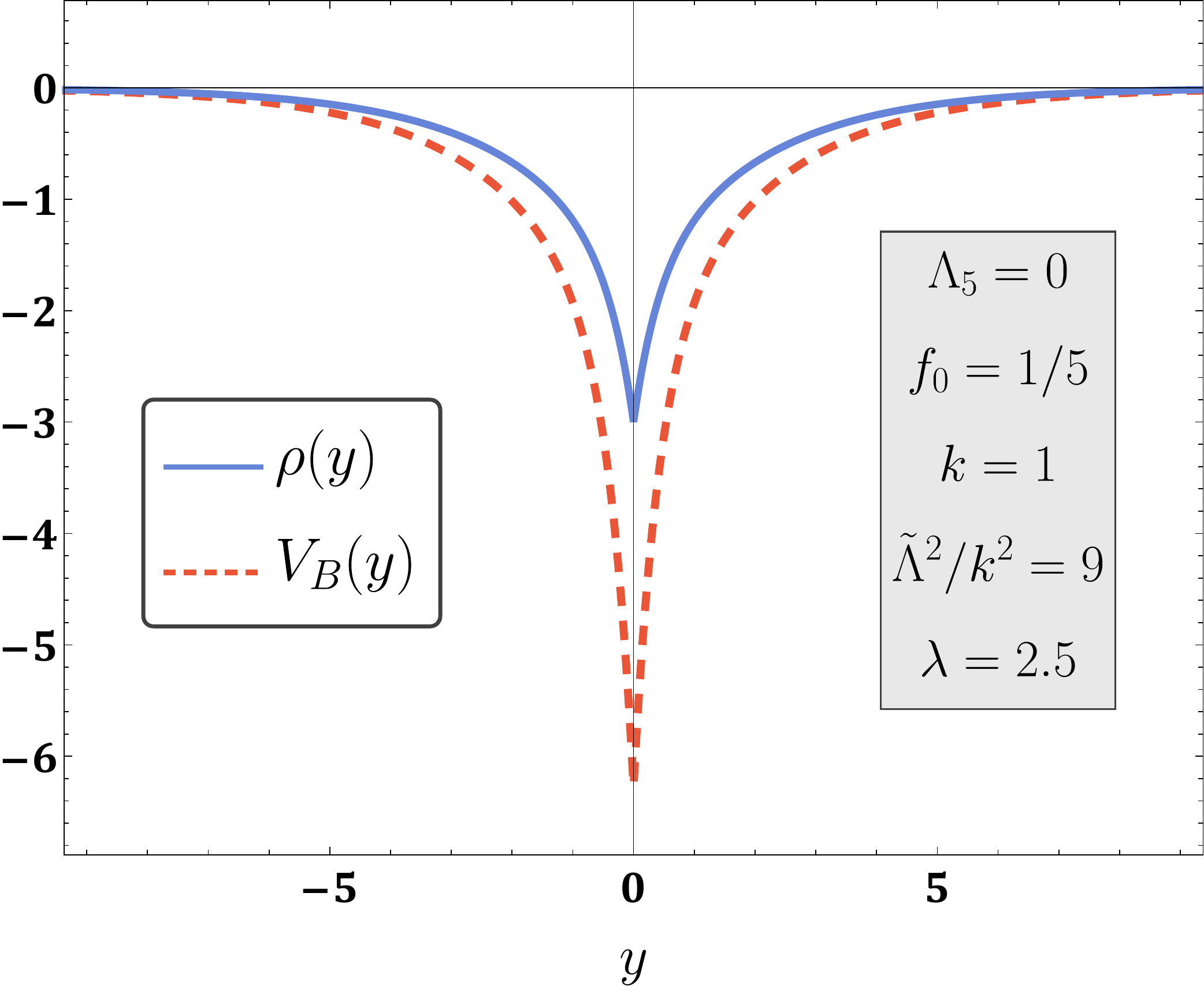}
        \caption*{\hspace{1.2em}(a)}
        \label{fig2a}
    \end{subfigure}
    ~ 
    \begin{subfigure}[b]{0.44\textwidth}
        \includegraphics[width=\textwidth]{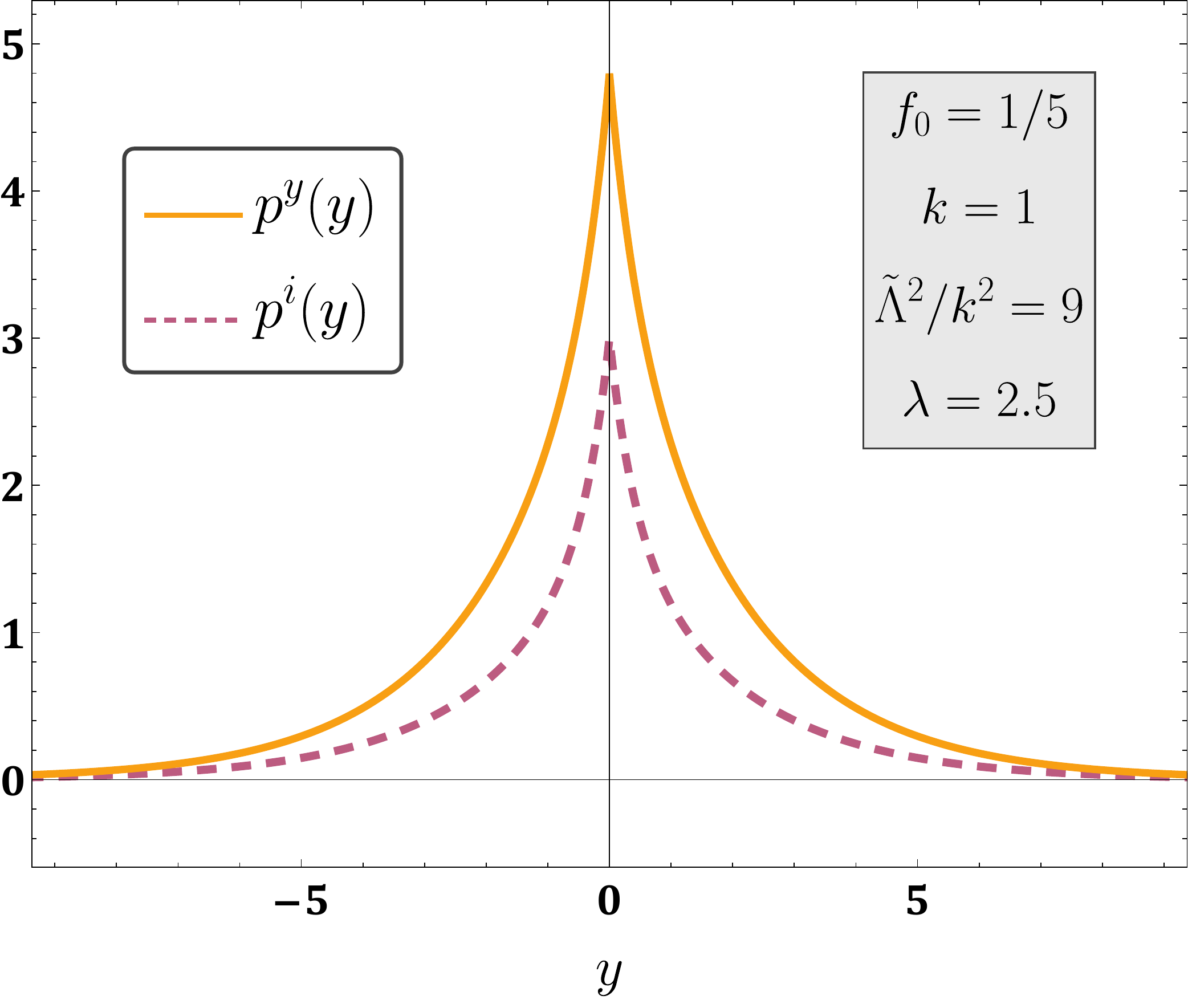}
        \caption*{\hspace{0.7em}(b)}
        \label{fig2b}
    \end{subfigure}
    ~ 
    \caption{(a) The scalar potential $V_B$ and energy density $\rho$ of the system, and
(b) the pressure components $p^y$ and $p^i$ in terms of the coordinate $y$.}
   \label{anti1-V-TMN}
\end{figure}

\par Finally, we may compute the components of the energy-momentum tensor of the theory 
in the bulk. These follow by employing Eqs. \eqref{Tmn} and \eqref{TMN-mixed}. Using also
the relations $\rho=-T^0{}_0$, $p^i=T^{i}{}_i$ and $p^y=T^y{}_y$, we find the results
\eq$\label{anti1-rho-0}
 \rho=-p^i=\frac{1}{2}\,\Phi'^2 + V_B+ \Lambda_5-3k \partial_y f +\partial_y^2 f\,,$
\eq$\label{anti1-p-0y}
p^y=\frac{1}{2}\,\Phi'^2 - V_B- \Lambda_5 + 4k \partial_y f\,.$
Substituting $V_B$ from Eq. (\ref{V_B-1}) and the form of the coupling function, we
finally obtain the following explicit expressions: 
\eq$\label{anti1-rho}
 \rho=-p^i=-f_0\,e^{-\lam ky}(6k^2+\tlam^2e^{2ky})\,,$
\eq$\label{anti1-py}
p^y=f_0\,e^{-\lam ky}(6k^2+2\tlam^2e^{2ky})\,.$
We present the behaviour of the energy density $\rho$ in Fig. 2(a) and of the pressure components
$p^i$ and $p^y$ in Fig. 2(b) with respect to the extra dimension $y$. Both figures have the same
values for the parameters of the model to allow for an easy comparison. The energy density is
negative-definite throughout the bulk, due to the negative value of the scalar potential discussed
above, in order to support the pseudo-AdS spacetime and the exponentially falling warp factor.
The spacelike pressure components $p^i$ satisfy the relation $p^i=-\rho$, a remnant of the
equation of state of a true cosmological constant. The fifth pressure component $p^y$ is also
positive but larger than $p^i$ due to the factor of 2 in front of $\tilde \Lambda^2$ 
in Eq. (\ref{anti1-py}). All components present a well-defined profile throughout the bulk and 
vanish exponentially fast away from the brane for all $\lambda>2$. The aforementioned behaviour
remains qualitatively the same for all values of the parameters of the model.


\subsection{Junction conditions and effective theory}

We will now turn our attention to the junction conditions that must be incorporated
in the model due to the presence of the brane at $y=0$. We will assume that the
energy content of the brane is given by the combination $\sigma + V_b(\Phi)$, 
where $\sigma$ is the constant self-energy of the brane and $V_b(\Phi)$ an interaction
term of the bulk scalar field with the brane. This energy content is assumed to arise
only at a single point along the extra dimension, i.e. along our brane at $y=0$, and
thus it creates a discontinuity
in the second derivatives of the warp factor, the coupling function and the scalar
field at the location of the brane. Following the standard literature \cite{BDL}, we write
$A''=\hat A'' + [A']\,\delta(y)$,  $f''=\hat f''+[f']\ \del(y)$ and $\Phi''=\hat \Phi''
+ [\Phi']\,\delta(y)$, where the hat quantities denote the distributional (i.e. regular)
parts of the second derivatives and $[\cdots]$ stand for the discontinuities of the
corresponding first derivatives across the brane. Then, going back to the field equations
(\ref{phi-eq}) and (\ref{grav-1}), we reintroduce the delta-function terms, that we
omitted while working in the bulk. If we then match the coefficients of the delta-function
terms appearing in Eqs. (\ref{phi-eq}) and (\ref{grav-1}), we obtain the following two
conditions
\gat$\label{jun_con1}
[\Phi'] = 4 [A']\,\partial_\Phi f + \partial_\Phi V_b\,,\\[3mm]
\label{jun_con2}
3 f(y) [A']= -[f']- (\sigma + V_b)\,,$
respectively, where all quantities are evaluated at $y=0^+$. Using the expressions for
the warp function $A(y)=-k |y|$ and the coupling function $f(y)=f_0 e^{-\lam k |y|}$
in Eq. (\ref{jun_con2}), and making use of the assumed $\mathbf{Z}_2$ symmetry in the
bulk, we readily obtain the constraint 
\gat$\label{anti1-jun1}
\sig+V_b(\Phi)\Big|_{y=0}=2kf_0(\lam+3)\,.$
We note that the combination of parameters on the r.h.s. of the above equation is
positive-definite, therefore, the total energy density of our brane is always positive.
The above constraint may be
used to determine the value of the warp-factor parameter $k$ in terms of the fundamental
quantities of the brane tension $\sigma$ and the scalar-field parameters ($f_0, \lam, V_b$).
We thus observe that, once we decide the form of the non-minimal coupling function, the
warping gets stronger the larger the interaction term of the scalar field with the brane is.

In order to evaluate the first constraint (\ref{jun_con1}), we write that:
$\partial_\Phi f= \partial_y f/\Phi'$ and $\partial_\Phi V_b= \partial_y V_b/\Phi'$.
We are allowed to do this since, as we showed in Sec. 2, the function $\Phi(y)$
does not possess any extrema in the bulk; therefore, $\Phi'(y)$ never vanishes. Then, 
multiplying both sides of Eq. (\ref{jun_con1}) by $\Phi'$ and using Eq. (\ref{anti-grav-1}),
we obtain the condition
\beq
\label{anti1-jun2.2}
\partial_y V_b\Bigr|_{y=0}=2f_0\left[\tlam^2-k^2\lam(\lam+3)\right].
\eeq
This second constraint may be used in a twofold way: for a non-trivial interaction
term $V_b$, it may serve to determine an independent parameter in its expression;
alternatively, under the condition that $V_b=const.$ and thus $\partial_y V_b=0$,
it may determine the value of the effective cosmological constant on the brane
to be $\Lambda=-\tilde \Lambda^2=-k^2 \lam (\lam+3)$, a value that is absolutely
compatible with the original constraint (\ref{anti1-con1}) that should hold on the brane.

Let us finally address the issue of the effective theory on the brane. For this, we need to
derive the four-dimensional effective action by integrating the complete five-dimensional
one $S=S_B+S_{br}$, over the fifth coordinate $y$. Before we proceed though, we present the
explicit forms of the five-dimensional curvature invariants whose general form for the
metric ansatz (\ref{metric}) is given in Appendix \ref{App-Invar}. Substituting the
mass function (\ref{mass-sol}) and the warp function $A(y)=-k|y|$ in these expressions,
we obtain:
\gat$ R = -20k^{2} + 4 \Lambda e^{2k|y|}\,, \nonumber\\[2mm]
R_{MN}R^{MN} = 80 k^{4} - 32k^{2}\Lambda e^{2k|y|} + 4\Lambda^{2} e^{4k|y|}\,,
\label{invar}\\[2mm]
R_{MNKL}R^{MNKL} = 40k^{4} - 16k^{2}\Lambda e^{2k|y|} + \frac{8\Lambda^{2}e^{4k|y|}}{3}
+\frac{48 M^2 e^{4k|y|}}{r^6}\,, \nonumber$
where $\Lambda=-\tlam^2$ is the negative constant appearing in the projected-on-the-brane
gravitational background (\ref{metric-brane}). The above expressions are valid in the domain
$y\in(-\infty,0)\cup(0,\infty)$, i.e. throughout the bulk where the second derivative of
the warp factor equals zero. Note that, by keeping $M$ in the form of the mass function
(\ref{mass-sol}), the obtained solutions describe clearly a black string with its singularity
at $r=0$ extending along the extra dimension. Setting, however, $M=0$, we obtain solutions
that are maximally symmetric on the brane and possess only a true singularity at the
boundary of the bulk spacetime where $y \rightarrow \infty$. 

This latter singularity was not present in the case of the black-string solution of Ref. \cite{CHR},
when $M=0$. We note that the singular terms in Eq. (\ref{invar}) are directly related to the
integration constant $\Lambda$ that, as we will soon see, will be interpreted as the
cosmological constant on the brane. Our scalar-tensor theory allows for solutions with
non-zero cosmological constant on the brane while the model employed in Ref. \cite{CHR}
assumed the Randall-Sundrum fine-tuning between the bulk cosmological constant and the
brane tension to ensure a flat brane. If we also set $\Lambda=0$ in our analysis, this
additional singularity disappears and we recover a regular AdS spacetime as in \cite{CHR}.  
However, we consider the presence of a non-zero cosmological constant on the brane
as an important feature of the solutions, the effect of which has not been adequately
studied in the literature. To this end, our analysis reveals that a non-zero $\Lambda$
on the brane is accompanied by a singularity in the bulk, located at an infinite coordinate
distance from our brane.\footnote{The presence of this singularity does not affect the
remaining features of the solution, such as the scalar field configuration, the warping
of spacetime, or the effective theory on the brane---however, if desired, it could easily
be shielded by the introduction of a second brane.} 

Returning to the four-dimensional effective theory, we will first calculate the effective
gravitational scale $M_{Pl}^2$ on the brane. To this end,
by employing the first of Eqs. (\ref{invar}), we may write $R=-20k^2 +R^{(4)} e^{2k|y|}$,
where $R^{(4)}=4 \Lambda$ is the four-dimensional scalar curvature, that may easily be
computed from the projected-on-the-brane line-element (\ref{metric-brane}). Hence, the term
from the complete action $S=S_B+S_{br}$ that is relevant for the evaluation of the effective
gravitational constant is the following:
\gat$ \label{action_eff}
S\supset\int d^4 x\,dy\,\sqrt{-g^{(5)}}\ \frac{f(\Phi)}{2}\ e^{2k|y|} R^{(4)}\,.$
Then, using also that $\sqrt{-g^{(5)}}=e^{-4k|y|} \sqrt{-g^{(br)}}$, where
$g^{(br)}_{\mu\nu}$ is the metric tensor of the projected on the brane spacetime,
the four-dimensional, effective gravitational constant is given by the integral
\eq$\frac{1}{\kappa_4^2}\equiv 2\,\int_{0}^{\infty} dy\, e^{-2 k y}\,f(y)
=2f_0\,\int_{0}^{\infty} dy\ e^{-2ky}e^{-\lam ky}=
\frac{2f_0}{k(\lam+2)}\,.
\label{anti1_effG}$
Since $1/\kappa_4^2=M_{Pl}^2/(8\pi)$, we obtain
\eq$
M_{Pl}^2=\frac{16 \pi f_0}{k(\lam+2)}=\frac{32 \pi f_0^2}{(\sigma+V_b)|_{y=0}}\,
\frac{(\lam +3)}{(\lam+2)}\,.$
In the last expression above, we have replaced the warp-factor parameter $k$
from Eq. (\ref{anti1-jun1}). We first note that the integral in Eq. (\ref{anti1_effG})
is finite; therefore, there is no need for the introduction of a second brane in the model
(unless one wishes to shield the singularity at the boundary of spacetime by introducing
a second brane). Also, according to the above result, the effective gravity scale
on the brane $M_{Pl}^2$ is determined by the ratio $f_0/k$. Taking into account that $f_0$
has units of $[M]^3$, after the absorption of $1/\kappa^2_5$ in its value, and thus
plays the role of the fundamental energy scale $M_*^3$, the relation between the
fundamental and the effective gravity scales turns out to be almost the same as the
corresponding one in the Randall-Sundrum II model \cite{RS2}. The difference between
the magnitudes of $M_*$ and $M_{Pl}$ is determined by the combination $f_0/(\sigma+V_b)|_{y=0}$
(note that $\lam$ plays virtually no role). Therefore, if a low-gravity scale is
desired, i.e. a low $f_0$, then the total energy density of the brane should be
minimized. This may be realised via the presence of a large, negative interaction
term $V_b$ for the scalar field that would result in a small, yet positive as dictated
by Eq. (\ref{anti1-jun1}), value for the combination $(\sigma+V_b)|_{y=0}$. 

To complete our study of the effective theory on the brane, we finally compute the
effective cosmological constant. Since the scalar field $\Phi$ is only $y$-dependent,
the effective theory contains no dynamical degree of freedom. Therefore, the
integral of all the remaining terms of the five-dimensional action $S=S_B+S_{br}$,
apart from the one appearing in Eq. (\ref{action_eff}), yields the effective
cosmological constant on the brane. Due to the existence of the brane, which acts
as a boundary for the five-dimensional spacetime, the bulk integral must be supplemented
by the source term of the brane as well as the Gibbons-Hawking term \cite{Gibbons-terms}.
In total, we have
\bea
\label{cosm_eff}
-\Lambda_4&=&\int_{-\infty}^{\infty} dy\,e^{-4k|y|}\Bigl[-10 k^2 f(y)-\Lambda_5-
\frac{1}{2}\,\Phi'^2 -V_B(y)+f(y)(-4A'')|_{y=0}-[\sigma +V_b(\Phi)]\,\delta(y)\Bigr]\nonum\\[1mm]
&=&2\int_0^\infty dy\ e^{-4ky}\left[-10 k^2 f(y)-\Lambda_5-
\frac{1}{2}\,\Phi'^2 -V_B(y)\right]+8kf(0)-[\sigma +V_b(\Phi)]_{y=0}\,.
\eea
Substituting the expressions for the coupling function and the bulk potential of the
scalar field, and employing the junction condition \eqref{anti1-jun1}, we finally
obtain the result
\eq$\label{anti1_effL}
\Lambda_4=-\frac{2f_0\ \tlam^2}{k(\lam+2)}=\frac{\Lambda}{\kappa_4^2}\,.$
As expected, the constant of integration $\Lambda$ appearing in the form of the mass
function (\ref{mass-sol}), and in the projected-on-the-brane line-element (\ref{metric-brane})
is indeed the four-dimensional cosmological constant $\Lambda_{4}$, multiplied by $\kappa_4^2$,
as the inverse Vaidya coordinate transformation on the brane had demonstrated \cite{KNP}. 


\section{The Double Exponential Case}
\label{anti2}

We now proceed to the alternative form of the non-minimal coupling function given in 
Eq. (\ref{anti-g2}). As in the previous section, the focus will be on the derivation
of analytic solutions of the field equations and the study of the characteristics of
the resulting solutions both in the bulk and on the brane.


\subsection{The bulk solution}

In this case, the coupling function has the form $f(y)=f_0\,e^{-\mu^2 e^{\lambda y}}$,
with $\lam$ being any positive real number and $\mu$ a real, non-vanishing number. Substituting
the function $g(y)=-\mu^2 e^{\lambda y}$ in the expression of the scalar field $\Phi(y)$
given by Eq. \eqref{anti-phi}, we obtain the integral expression
\eq$\label{anti2-phi}
\Phi_{\pm}(y)=\pm \sqrt{f_0}\int dy \exp\left(-\frac{\mu^2}{2}e^{\lam y}\right)
\sqrt{\tlam^2 e^{2ky}+\mu^2\lam(\lam+k)e^{\lam y}-\mu^4\lam^2e^{2\lam y}}\,.$
In general, the above integral does not have an analytic solution. However, if one chooses
appropriate values for the parameters $\mu^2$ and $\lam$, the quantity under the square root
can be expressed as a perfect square and the integral becomes solvable. To this end,
we can rewrite the aforementioned quantity as:
\gat$\label{per-sqrt}
\tlam^2 e^{2ky}+\mu^2\lam(\lam+k)e^{\lam y}-\mu^4\lam^2e^{2\lam y}=
\left(\sqrt{\tlam^2}\,e^{ky}-\sqrt{\mu^2\lam(\lam+k)}\,e^{\lam y/2}\right)^2\,,$
provided that the following conditions are imposed:
\eq$\label{anti2-cons}
k+\frac{\lam}{2}=2\lam\,, \qquad 2\sqrt{\tlam^2\mu^2\lam(\lam+k)}=\mu^4\lam^2\,.$
These lead to the unique values 
\beq
\lam=\frac{2k}{3}\,, \qquad \mu^2=\left(\frac{45}{2}\frac{\tlam^2}{k^2}\right)^{\frac{1}{3}}\,,
\label{lam-mu}
\eeq
for the $\lam$ and $\mu$ parameters. Using the form of the coupling function in
Eq. \eqref{anti-grav-1}, and substituting the above values for $\lam$ and $\mu^2$ 
in the result, we are led to
\gat$ \label{anti2-Phi/f}
\Phi'^2(y)=f(y)\,\Bigl[\tlam^2 e^{2ky}+\mu^2\lam(\lam+k)e^{\lam y}-
\mu^4\lam^2e^{2\lam y}\Bigr]=
\frac{2k^2\mu^2}{45}\,f(y)\,e^{\frac{2ky}{3}}\left(\mu^2e^{\frac{2ky}{3}}-5\right)^2\,.$
The above equation can provide important information on the form of the scalar
field in the bulk even before the explicit integration in Eq. ({\ref{anti2-phi}) is
performed. To start with, since $f(y)>0$ for all $y>0$, the r.h.s. of the above relation
is automatically positive-definite; therefore, no additional constraint on the parameters
of the theory follows by demanding the positivity of $\Phi'^2$.
The value of the parameter $\mu^2$ though affects significantly the profile of the scalar field
along the extra dimension. In particular, if the value of $\mu^2$ is lower than 5, then the
first derivative of the scalar field will become zero at 
$y_0=\frac{3}{2k}\ln\left(\frac{5}{\mu^2}\right)$. If the value of $\mu^2$ is exactly 5,
then the first derivative of the
scalar field is zero at $y=0$. Finally, if $\mu^2$ is greater than 5, then the first
derivative of the scalar field does not vanish anywhere in the bulk. In summary,
\eq$\label{val-for-mu}
\left\{\begin{array}{l}
\mu^2<5: \hspace{1.5em}\Phi'(y_0)=0,\hspace{1.5em}\displaystyle{y_0=\frac{3}{2k}\ln\left(\frac{5}{\mu^2}\right),}\\[4mm]
\mu^2=5: \hspace{1.5em}\Phi'(0)=0,\\[4mm]
\mu^2>5: \hspace{1.5em}\Phi'(y)\neq 0, \hspace{2.3em}\forall y>0.\end{array}\right\}$
Taking however the square root of Eq. (\ref{anti2-Phi/f}), we obtain for $\Phi'(y)$
the expression
\beq
\Phi'(y)=\pm k \sqrt{\frac{2 f_0 \mu^2}{45}}\,\exp\left(-\frac{\mu^2}{2}\,e^{\frac{2ky}{3}}
+\frac{ky}{3}\right)\,\left|\mu^2e^{\frac{2ky}{3}}-5\right|.
\label{anti2-phi'}
\eeq
We thus conclude that $\Phi'(y)$ retains a specific sign throughout the bulk, even in
the case where $\mu^2<5$; therefore, the point $y=y_0$ is not an extremum
[where $\Phi'(y)$ owes to change its sign] but rather an inflection point. As a
result, $\Phi(y)$ is a monotonic function throughout the bulk, for all values of $\mu^2$,
and thus it is a one-to-one function in the whole region $y>0$ (as well as in the $y<0$
region due to the ${\mathbf{Z_2}}$ symmetry).

The above behaviour also affects the way that one should proceed in order to find
the solution for the scalar field $\Phi$. For $\mu^2\geq 5$, the quantity inside
the absolute value in Eq. (\ref{anti2-phi'}) is positive and non-vanishing for all
values $y>0$; thus, the solution for $\Phi(y)$, at every point in the bulk, follows
by directly integrating Eq. (\ref{anti2-phi'}). Then, we obtain
\gat$
\Phi_{\pm}(y)=\pm \left[\mathcal{I}(y)-\mathcal{I}(0)\right],$
where we have defined $\mathcal{I}(y)$ as 
\eq$ \label{I-function}
\mathcal{I}(y)\equiv -\sqrt{\frac{f_0}{5}}\left[\sqrt{2\mu^2} \exp\left(\frac{k y}{3}-\frac{1}{2}
\mu^2 e^{\frac{2 k y}{3}}\right)+
4 \sqrt{\pi}\ \text{erf}\left(\sqrt{\frac{\mu^2}{2}}e^{\frac{k y}{3}}\right)\right],$
and $\text{erf}(z)$ is the error function
\beq
\text{erf}(z)=\frac{2}{\sqrt{\pi}}\int_0^z e^{-t^2}dt=\frac{2}{\sqrt{\pi}}
\sum_{n=0}^\infty \frac{(-1)^n\,z^{2n+1}}{n!\,(2n+1)}\,.
\label{error}
\eeq
On the other hand, for $\mu^2<5$, we need to address separately the cases where the
solution for $\Phi(y)$ is found at a point in the bulk with $y\leq y_0$ or at a point
beyond the inflection point with $y > y_0$. In the first case, apart from the change
in the order of the terms inside the absolute value in Eq. (\ref{anti2-phi'}), no
other action is necessary, and the integration over $y$ is performed as before. In
the second case, however, care must be taken when the inflection point at $y=y_0$
is reached. Then, we write:
\bal$
\Phi_{\pm}(y)&=\pm k \sqrt{\frac{2 f_0 \mu^2}{45}}\,\left[\int_0^{y_0} dy' 
\exp\left(-\frac{\mu^2}{2}e^{\frac{2ky'}{3}}+\frac{ky'}{3}\right)
\left(5-\mu^2e^{\frac{2ky'}{3}}\right)\right.\nonum\\[3mm]
&\hspace{2.5em}\left.+\int_{y_0}^y dy' \exp\left(-\frac{\mu^2}{2}e^{\frac{2ky'}{3}}+
\frac{ky'}{3}\right)\left(\mu^2e^{\frac{2ky'}{3}}-5\right)\right].$
Overall, for $\mu^2<5$, the solution for the scalar field is
\eq$\label{anti2-phi-mu<}
\Phi_{\pm}(y)=\left\{\begin{array}{cr}
\mp\left[\mathcal{I}(y)-\mathcal{I}(0)\right],&\hspace{1em} y\leq y_0,\\[4mm]
\pm\left[\mathcal{I}(0)-2\mathcal{I}(y_0)+\mathcal{I}(y)\right],& y > y_0,\\[4mm]
\end{array}\right\}$
where $\mathcal{I}(y)$ is still given by Eq. (\ref{I-function}).

\begin{figure}[b!]
    \centering
    \begin{subfigure}[b]{0.46\textwidth}
        \includegraphics[width=\textwidth]{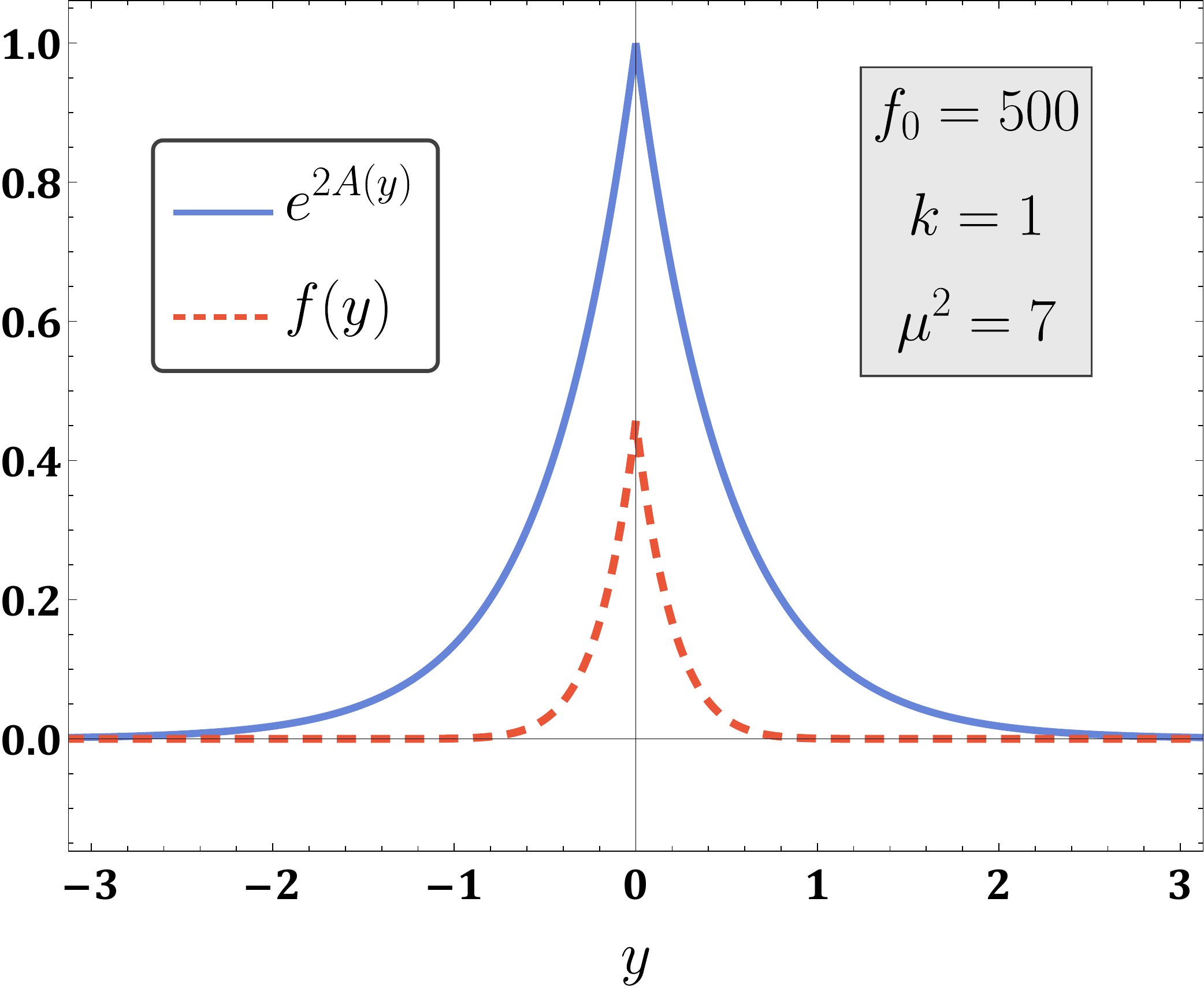}
        \caption*{\hspace{1em}(a)}
        \label{fig3a}
    \end{subfigure}
    ~ 
    \begin{subfigure}[b]{0.49\textwidth}
        \includegraphics[width=\textwidth]{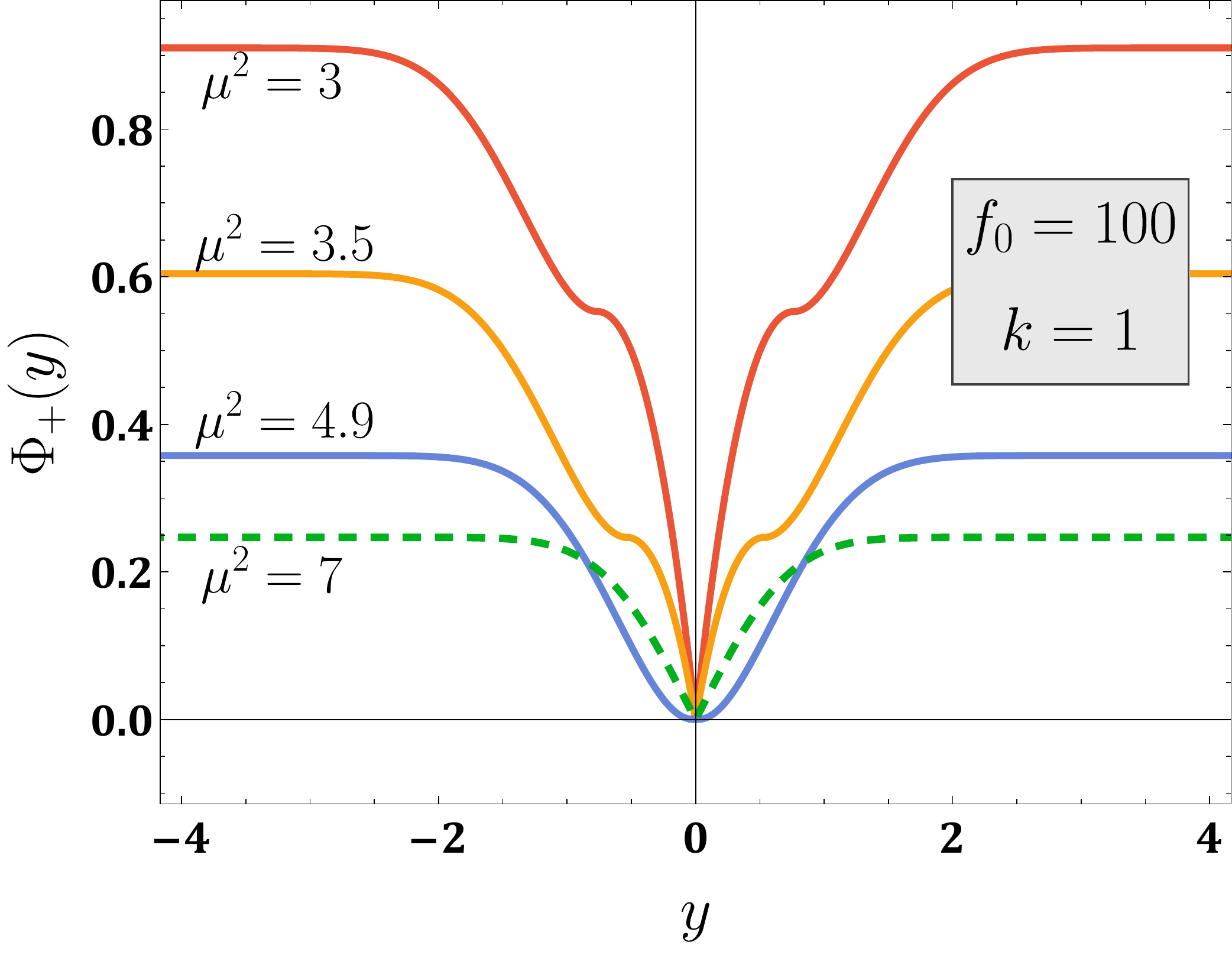}
        \caption*{\hspace{2.8em}(b)}
        \label{fig3b}
    \end{subfigure}
    ~ 
    \caption{(a) The warp factor $e^{2A(y)}=e^{-2k|y|}$ and the coupling function
    $f(y)=f_0\,e^{-\mu^2 e^{2k y/3}}$, and (b) the scalar field $\Phi_{+}(y)$ for
    various values of the parameter $\mu^2$, in terms of the coordinate $y$.}
   \label{anti2_warp_f_Phi}
\end{figure}

\par In Fig. \ref{anti2_warp_f_Phi}(a), we depict the form of the warp factor $e^{-2k|y|}$
and the coupling function $f(y)=f_0\,e^{-\mu^2 e^{2k y/3}}$ in terms of the coordinate
$y$ along the fifth dimension, for $f_0=500$, $k=1$ and $\mu^2=7$. Both functions
exhibit a localization close to the brane with the coupling function $f(y)$ decreasing,
in fact, much faster due to its double exponential dependence on $y$. At the boundary of
spacetime, both functions go smoothly to zero. The displayed, qualitative behaviour of
these two quantities is independent of the particular values of the parameters.
In contrast, the profile of the scalar field $\Phi(y)$ with respect to the extra dimension
$y$ depends strongly on the value of the parameter $\mu^2$, as one may clearly see
in Fig. \ref{anti2_warp_f_Phi}(b). We observe that the behaviour of the scalar field
changes significantly as the parameter $\mu^2$ approaches and then surpasses the value 5.
Indeed, for $\mu^2 <5$, the emergence of the  inflection point at $y=y_0>0$ is clearly
visible. As $\mu^2$ approaches the value 5, the inflection point moves towards the brane.
For $\mu^2 \geq 5$, though, this feature completely disappears in accordance with the
analytical study presented above. Overall, the scalar field exhibits a monotonic behaviour
over the entire bulk---for the $\Phi_+(y)$ solution that we have chosen here to plot, the scalar
field presents an increasing profile in the bulk reaching a constant, asymptotic value at
the boundary of spacetime.  In Figs. \ref{anti2_Phi_2_3}(a) and \ref{anti2_Phi_2_3}(b),
we present the dependence
of $\Phi_+(y)$ on the second parameter $k$, since the parameter $\lam$ is now fixed to
the value of the warping parameter $k$ through the first of Eqs. (\ref{lam-mu}); henceforth,
we drop any reference to $\lam$. We observe again the emergence of the inflection point
when $\mu^2 <5$, in Fig. \ref{anti2_Phi_2_3}(a), and the smooth behaviour when $\mu^2 > 5$,
in Fig. \ref{anti2_Phi_2_3}(b). The value of the warping parameter $k$ causes only a
rise in the slope of the curve, as $k$ increases, leaving all the other features invariant.

\begin{figure}[t!]
    \centering
    \begin{subfigure}[b]{0.485\textwidth}
        \includegraphics[width=\textwidth]{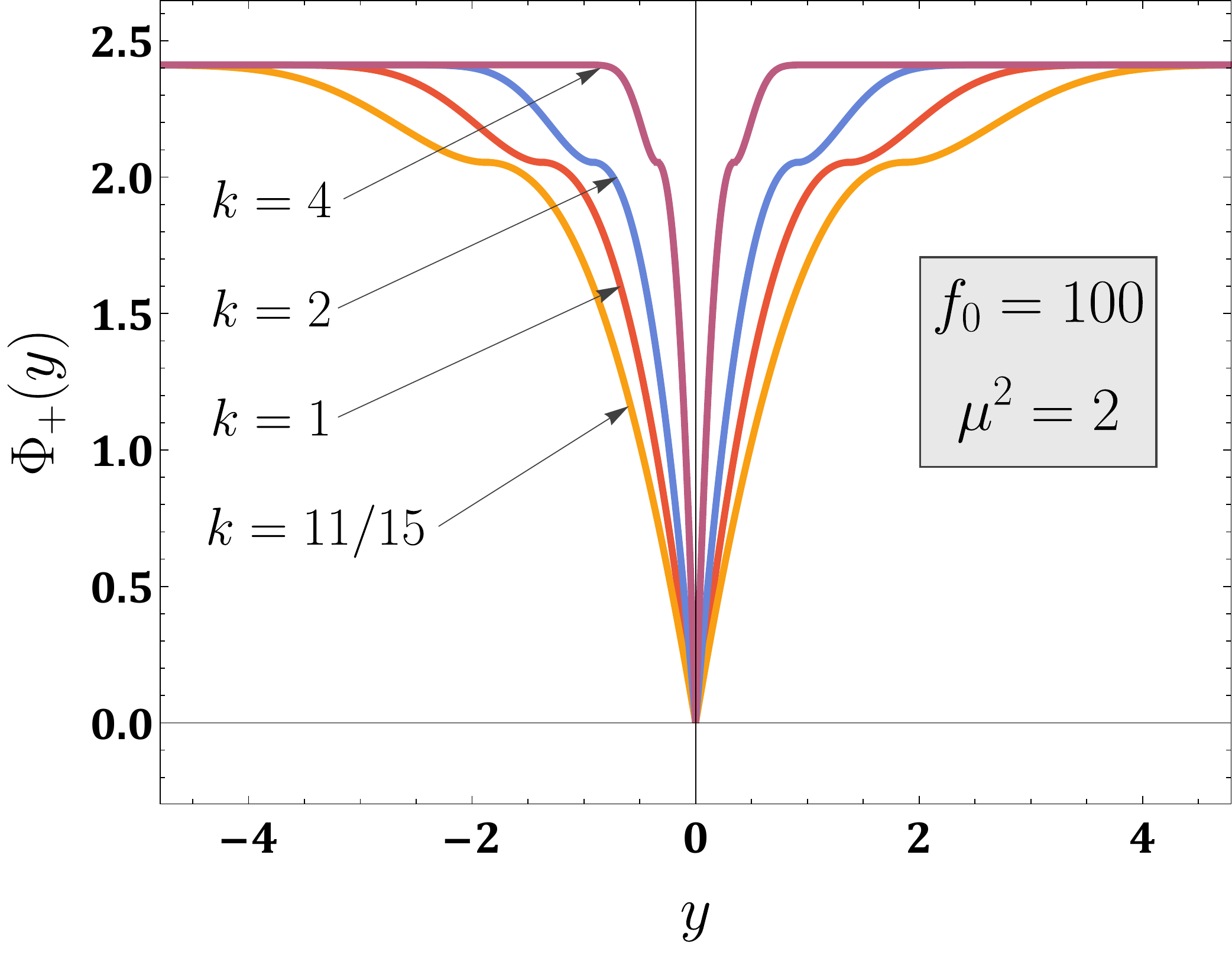}
        \caption*{\hspace{2.9em}(a)}
        \label{fig4a}
    \end{subfigure}
    ~ 
    \begin{subfigure}[b]{0.49\textwidth}
        \includegraphics[width=\textwidth]{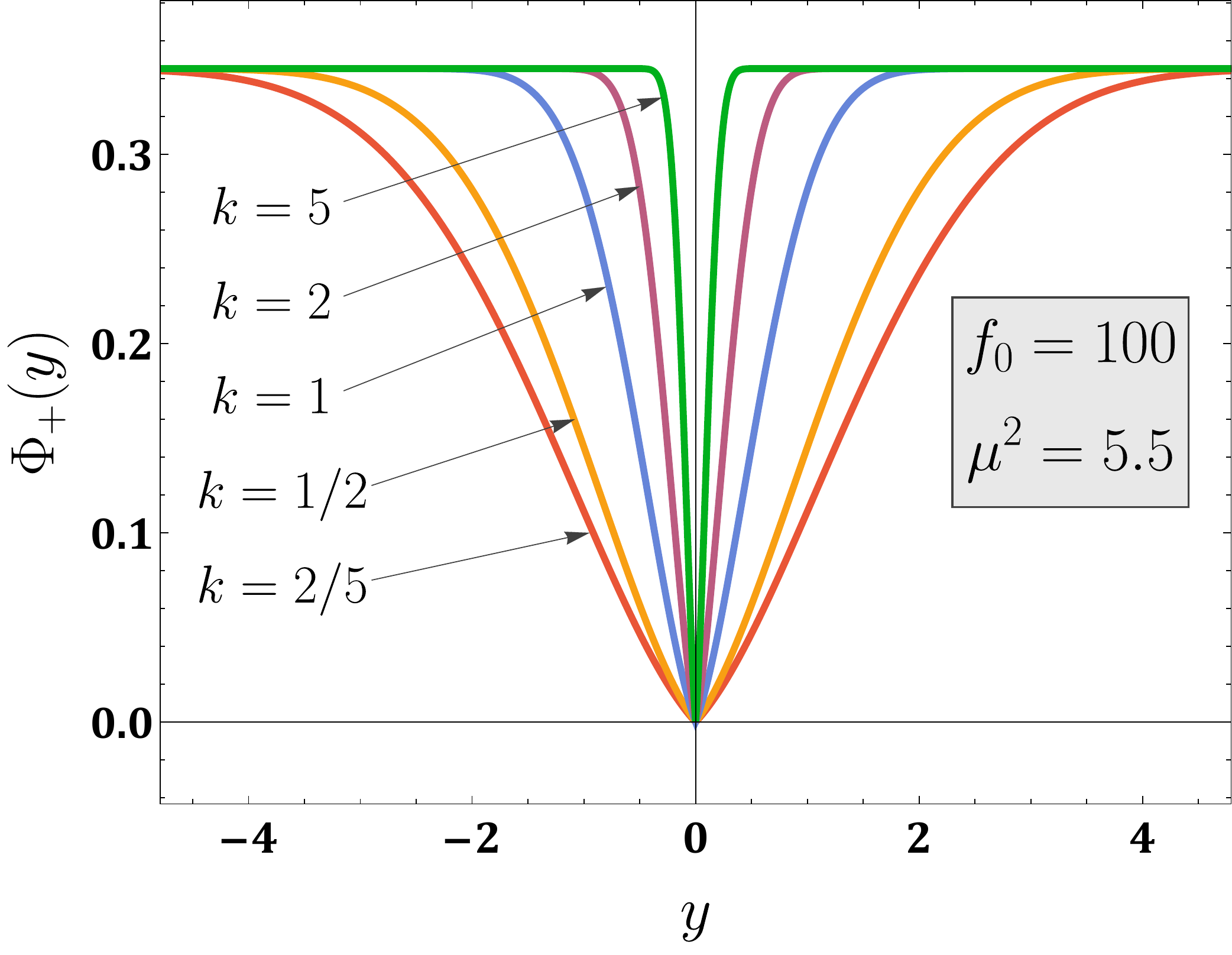}
        \caption*{\hspace{2.9em}(b)}
        \label{fig4b}
    \end{subfigure}
    ~ 
    \caption{The scalar field $\Phi_{+}(y)$, in terms of the coordinate $y$, for $f_0=100$ and
    for (a) $\mu^2=2$, and (b) $\mu^2=5.5$ and various values of $k$. }
   \label{anti2_Phi_2_3}
\end{figure}

\par The bulk potential of the field $V_B$ in this case can be determined again by the
general expression (\ref{V_B-1}). Substituting the double exponential form of the coupling
function $f(y)$, we now obtain the result
\eq$\label{anti2-pot}
V_B(y)=-\Lambda_5-f_0\ e^{-\mu^2e^{\frac{2ky}{3}}}\left[6k^2+\frac{k^2\mu^2}{45}
e^{\frac{2ky}{3}}\left(3\mu^4e^{\frac{4ky}{3}}+10\mu^2e^{\frac{2ky}{3}}+95\right)\right].$
As in the simple exponential case, the combination inside the square brackets
in the above expression is positive-definite, for all values of the parameters of the model,
thus rendering the second term of the bulk potential negative-definite. Therefore, the
presence of a non-minimally-coupled scalar field in the bulk leads to a negative (non-constant) 
potential energy in the bulk that can support again an AdS-type bulk
spacetime with an exponentially decreasing warp factor, even if the quantity $\Lambda_5$ is
set to zero. The potential has a smooth form over the entire bulk, is localized close
to the brane and goes to zero extremely fast away from it---all these features are
inherited from the form of the coupling function to which $V_B$ is directly proportional
as  Eq. (\ref{anti2-pot}) clearly shows. The aforementioned behaviour of $V_B$ is depicted
in Fig. \ref{anti2_V-TMN}(a) for $f_0=1/5$, $k=1$ and $\mu^2=5$.

\par The components of the energy-momentum tensor of the theory may be computed employing
again Eqs. (\ref{anti1-rho-0}) and (\ref{anti1-p-0y}). Substituting again the form of
the coupling function together with the expression for the bulk potential (\ref{anti2-pot})
presented above, we find the explicit expressions
\eq$\label{anti1-rho-2}
\rho=-p^i=-2f_0k^2\left(3+\frac{\mu^6}{45}e^{2ky}\right)e^{-\mu^2e^{\frac{2ky}{3}}}\,,$
\eq$\label{anti1-py-2}
p^y=2f_0k^2\left(3+\frac{2\mu^6}{45}e^{2ky}\right)e^{-\mu^2e^{\frac{2ky}{3}}}\,.$
In Figs. \ref{anti2_V-TMN}(a) and \ref{anti2_V-TMN}(b) we present the behaviour of the energy
density $\rho$ and the pressure components $p^i$ and $p^y$, respectively, in terms of the
extra dimension $y$. These quantities, too, present a smooth profile over the entire bulk,
remain localised
close to our brane, and vanish asymptotically leaving behind a 5-dimensional, flat spacetime
(if $\Lambda_5$ is assumed zero). The energy density $\rho$ is again negative throughout the
bulk (as it should be in order to support by itself a pseudo-AdS spacetime) but this is due to 
the presence of a physical, scalar degree of freedom coupled non-minimally to gravity with
a physically-acceptable positive-definite, and localised close to our brane, coupling
function.  

\begin{figure}[t!]
    \centering
    \begin{subfigure}[b]{0.485\textwidth}
        \includegraphics[width=\textwidth]{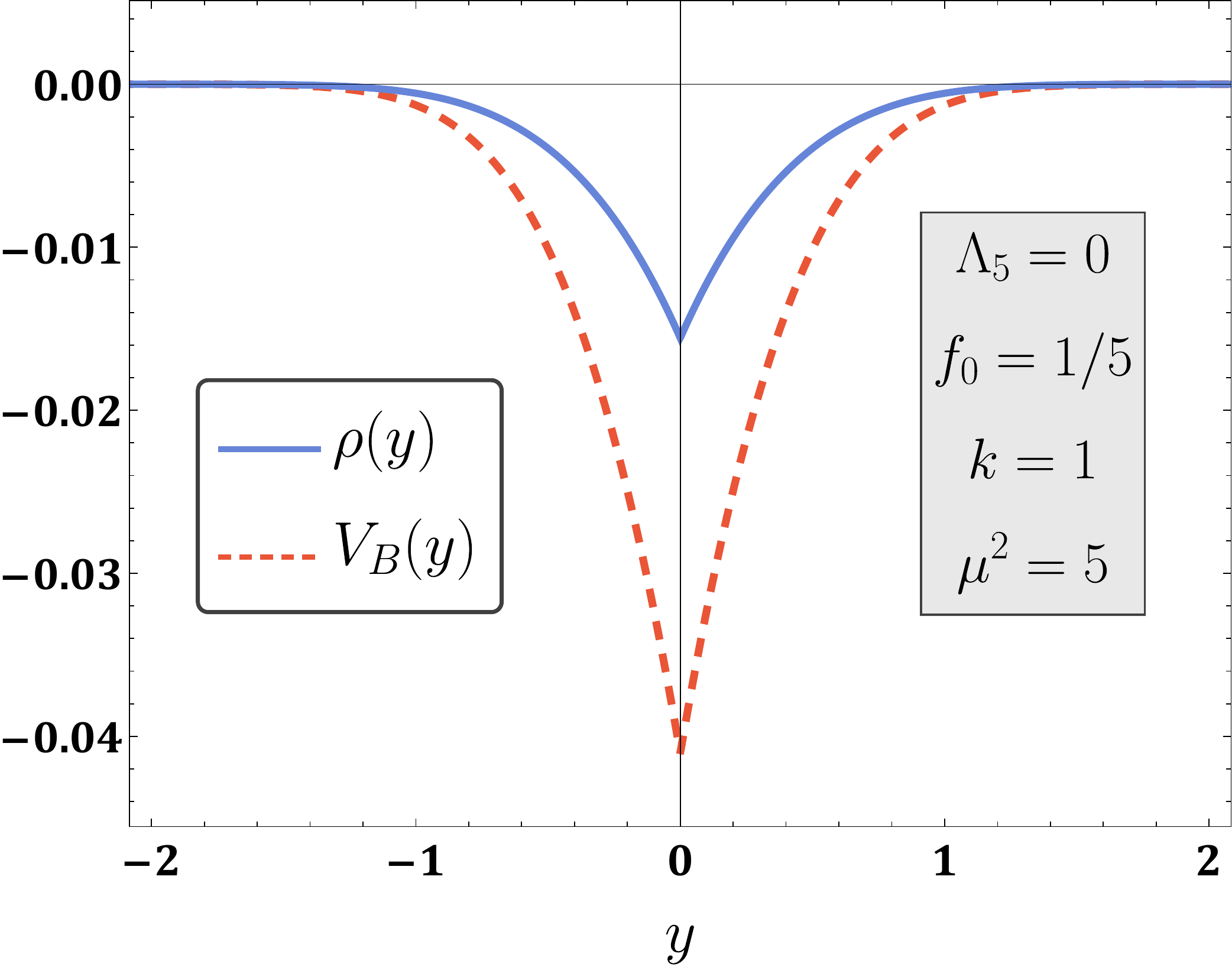}
        \caption*{\hspace{2.2em}(a)}
        \label{fig5a}
    \end{subfigure}
    ~ 
    \begin{subfigure}[b]{0.49\textwidth}
        \includegraphics[width=\textwidth]{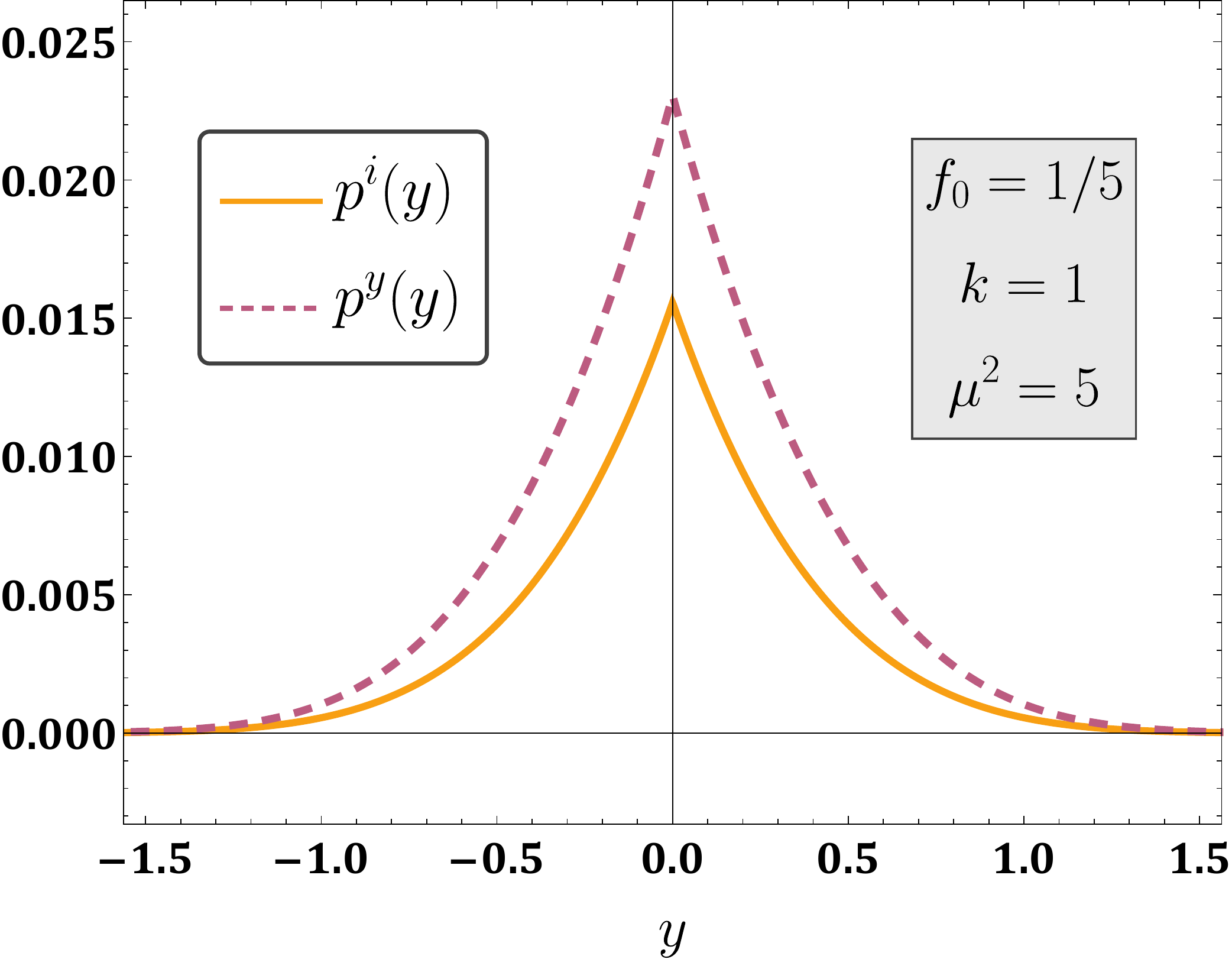}
        \caption*{\hspace{2em}(b)}
        \label{fig5b}
    \end{subfigure}
    ~ 
    \caption{(a) The scalar potential $V_B$ and energy density $\rho$ of the system, and
(b) the pressure components $p^y$ and $p^i$ in terms of the coordinate $y$. }
   \label{anti2_V-TMN}
\end{figure}


\subsection{Junction conditions and effective theory}

For the junction conditions, we use again the general expressions \eqref{jun_con1} and
\eqref{jun_con2}, in which we substitute the form of the coupling function and employ
also Eq. (\ref{anti2-phi'}) to replace $\Phi'(y)$. Then, Eq. \eqref{jun_con2}
straightforwardly leads to the constraint
\gat$\label{anti2-1-jun1}
2kf_0e^{-\mu^2}\left(3+\frac{2\mu^2}{3}\right)=\sig+V_b(\Phi)|_{y=0}\,,$
which may be used again to fix the warping parameter $k$ in terms of the 
parameters $f_0$ and $\mu$ of the coupling function and the energy content
$(\sigma+V_b)$ of the brane. On the other hand, Eq. \eqref{jun_con1} 
results\footnote{Note that care should be taken in the
evaluation of \eqref{jun_con1} due to the different behaviour of the scalar field
$\Phi(y)$ in terms of $\mu^2$. At the end of the evaluation, though, a unique expression
follows from this junction condition for either $\mu^2>5$ or $\mu^2<5$.} in the condition
\gat$
\label{anti1-1-jun2.2}
V'_b(0)=f_0\ e^{-\mu^2}\frac{4\mu^2k^2}{3}\left[\frac{(\mu^2-5)^2}{15}-4\right].$
For a non-trivial $V_b$, the above condition can be used to restrict a parameter that
may appear in its expression; on the other hand, for a trivial $V_b$, setting
$V'_b(0)=0$ in Eq. (\ref{anti1-1-jun2.2}}), we may fix also the value of $\mu^2$,
and, through Eq. (\ref{lam-mu}), the value of the effective cosmological constant
on the brane $\Lambda=-\tilde \Lambda^2$.

We turn finally to the 4-dimensional, effective theory on the brane. In order to
derive the effective gravitational constant on the brane, we may use again the
relation \eqref{action_eff}. Substituting the double exponential form of the
coupling function $f(y)$, we find:
\bal$\label{anti2-kappa4}
\frac{1}{\kappa_4^2}&=2f_0\int_{0}^{\infty} dy\ e^{-2ky}\exp\left(-\mu^2
e^{\frac{2ky}{3}}\right)=\frac{f_0}{2k}\left[e^{-\mu^2}\left(2-\mu^2+\mu^4\right)+
\mu^6\ \text{Ei}\left(-\mu^2\right)\right].$
Above, we have used the exponential-integral function $\text{Ei}(x)$ defined as
\beq
\text{Ei}(x)\equiv-\int_{-x}^\infty dt\ \frac{e^{-t}}{t}\,,
\label{fun-Ei}
\eeq
and its property that $\lim_{x\ra\infty}\text{Ei}(-x)=0$. For $x<0$, it can also be shown that
\cite{Abramowitz}
\beq
\text{Ei}(x)=\text{Ei}\left(-|x|\right)=\gamma+\ln|x|+\sum_{k=1}^\infty 
\frac{(-1)^{k}|x|^k}{k\ k!}\,,
\eeq
where $\gamma$ is the Euler-Mascheroni constant. Then, the effective gravitational
energy scale is given by
\bal$\label{anti2-MPl}
M_{Pl}^2 &=\frac{4\pi f_0}{k}\left\{e^{-\mu^2}\left(2-\mu^2+\mu^4\right)+\mu^6\left[\gamma+
2\ln|\mu|+\sum_{q=1}^\infty \frac{(-1)^{q}\mu^{2q}}{q\ q!}\right]\right\}.$
Once again, the gravity scale on the brane $M_{Pl^2}$ is determined primarily by the
ratio $f_0/k \sim M_*^3/k$, which resembles again the corresponding relation of the
Randall-Sundrum model \cite{RS2}---note that the combination inside the curly brackets
in Eq. (\ref{anti2-kappa4}) is of ${\cal O}(1)$. 

The effective cosmological constant on the brane can be evaluated by employing
Eq. \eqref{cosm_eff}. Using again the form of the coupling function $f(y)$ and the
junction condition (\ref{anti2-1-jun1}), we find that
\gat$
\Lambda_4=-\frac{f_0 k\mu^6}{45}\left[e^{-\mu^2}\left(2-\mu^2+\mu^4\right)+
\mu^6\,\text{Ei}\left(-\mu^2\right)\right]=
-\frac{\tlam^2}{\kappa_4^2}=\frac{\Lambda}{\kappa_4^2}\,.$
Above, we have used the second of Eqs. (\ref{lam-mu}), which relates the parameter
$\mu^2$ with $\tilde \Lambda^2$, and the expression (\ref{anti2-kappa4}) for $\kappa_4^2$.
As in the model discussed in Sec. 3, the integration parameter $\Lambda$ is confirmed
to be the effective cosmological constant $\Lambda_4$ on the brane multiplied by the
effective gravitational constant $\kappa_4^2$.

\section{Conclusions}

In this work, we have considered a five-dimensional gravitational theory containing 
a scalar field with a non-minimal coupling to the five-dimensional Ricci scalar. 
The coupling is realised through a smooth, real, positive-definite coupling function
$f(\Phi)$. Demanding that all components of the energy-momentum tensor remain
finite throughout the bulk, and looking for analytic solutions for the scalar
field, we have restricted our choices for the coupling function to two particular
forms: a simple exponential and a double exponential, both decreasing
away from the brane. This results into a scalar-tensor five-dimensional
theory with a non-minimal coupling between the scalar field and gravity that is
effectively localised close to the brane. 

The five-dimensional line-element contained two, initially arbitrary, functions: the warp
factor and the mass function. The warp factor was chosen to have the Randall-Sundrum
form in order to ensure the gravity localisation close to our brane. The mass function
was uniquely determined by integrating the gravitational field equations and turned
out to have a form that, on the brane, produces a Schwarzschild-(anti)-de Sitter
spacetime. Having studied the case of a positive effective cosmological constant
on the brane in a previous work of ours \cite{KNP}, here, we focused on the case
of a negative four-dimensional cosmological constant. In fact, for this choice,
the non-minimal coupling function is allowed to have everywhere a positive value
and thus to guarantee a normal gravitational force both in the bulk and on the brane.

The solution for the mass function contained also a mass parameter that was
identified with the mass of the black hole that a four-dimensional observer sees
on the brane. However, the calculation of the five-dimensional scalar-invariant
quantities revealed that, for a non-vanishing mass parameter, the five-dimensional
solution is in fact a black-string plagued by a singularity extending over the
entire extra dimension. It is again remarkable how easily the classical, ill-defined
black-string solutions emerge in the context of physically plausible scalar field
theories whereas classical black-hole solutions are still eluding us. If we set
this mass parameter equal to zero, then the extended singularity disappears
leaving behind a maximally symmetric spacetime on the brane and a regular
five-dimensional bulk with a true singularity at the boundary of spacetime.

In order to complete the solution, the form of the scalar field in the bulk also
had to be determined. We have managed to analytically attack the problem of
the integration of the scalar field equation and to derive two particular
solutions. For
the exponentially decreasing coupling function, the scalar field was expressed
in terms of the hypergeometric function (which, for particular values of the
parameters, was further reduced to a combination of elementary functions) while
for the double exponential form of $f(y)$, the scalar field was given by a combination
of an exponential and the error function. In both cases, the profile of the
scalar field presented the same set of robust features: $\Phi(y)$ was everywhere a
smooth, regular, monotonic function of the extra coordinate approaching a constant,
finite value at the boundaries of spacetime. 

The same robust behaviour was exhibited also by the bulk potential of the scalar
field and all components of the energy-momentum tensor. All the aforementioned
quantities were finite everywhere in the bulk, remained localised close to our
brane, and vanished asymptotically at large distances. In particular, the scalar
potential was everywhere negative-definite, which led also to a negative-definite
energy density. However, this negative distribution of energy was generated, not
by an exotic form of matter, but by a physical, scalar degree of freedom coupled
non-minimally to gravity with a positive-definite, and localised close to our brane,
coupling function. This energy may therefore support by itself a pseudo-AdS spacetime,
even in the absence of a negative, bulk cosmological constant, and thus ensure
the localisation of gravity close to our brane. 

The presence of the brane in the theory introduces a set of junction conditions
that may serve to fix two of the parameters of the theory, preferably the
warp-factor parameter $k$ and a parameter of the interaction term of the
scalar field with the brane. If the latter term is non-trivial, the effective
cosmological constant remains a free parameter of the model; however, for a
trivial scalar-brane interaction term, the value of the cosmological constant on
the brane may be uniquely determined. The calculation of the effective theory
on the brane led to a finite theory, as expected, without the need to introduce
a second brane in the model. The relation between the fundamental and the
effective gravitational scale had a similar form with the one emerging in the
Randall-Sundrum model although our theory has a dynamical, more realistic
field content.  

In order to produce finite, analytic solutions for the bulk scalar field, we have
made indeed two particular choices for the non-minimal coupling function.
However, despite the apparently different explicit forms of the scalar field,
the main characteristics of the two solutions remained the same, namely, the
smoothness, the regularity, even the constant, finite value at infinity. This
type of ``universality'' is caused by the common characteristics that the
corresponding forms of the coupling function had: they were both smooth,
well-defined throughout the bulk, positive-definite and localised close to
our brane. We may therefore assert that the two particular solutions we have
derived are in fact representative of the behaviour that a generic solution for the
scalar field would exhibit if it was sourced by the Ricci scalar through any
coupling function that would respect the same finiteness and positivity-of-value
criteria.

Let us stress again that the solutions presented in this work are supported by
a bulk distribution of energy related to a physically acceptable, realistic
scalar-tensor theory of gravity. Both of our solutions are characterised by
an exponentially decreasing warp factor, identical to that of Refs. \cite{RS1, RS2}, 
without assuming that the sole energy content of the theory in the bulk is
a negative cosmological constant---rather, it is a non-trivial scalar field
with a localised, negative potential that supports this feature. Our solutions
reduce to a black-hole line-element on the brane, as in Ref. \cite{CHR},
if we assume that $M \neq 0$, but allow also for the presence of a non-zero
cosmological constant on the brane. This feature was absent in the analysis
of \cite{CHR} where the brane was assumed to be flat. Allowing for a non-zero
cosmological constant on the brane has also revealed that the bulk possesses
then an additional singularity located at an infinite coordinate distance from our
brane in the case where $M = 0$. This singularity does not affect either
the profile of the scalar field in the bulk or the effective theory on the brane,
but, if desired, it could easily be shielded by the introduction of a second brane. 

Finally, the stability behaviour of our solutions is an important aspect that needs
to be studied. Compared to the existing stability analyses performed along the lines
of Refs. \cite{GL,RuthGL}, our theory has the additional complexity of the presence of
the scalar field. Our solutions are not purely gravitational; therefore, the perturbation
analysis will involve a coupled system of scalar-field and gravitational equations.
The sign of the cosmological constant $\Lambda$ and the corresponding properties
of the non-minimal coupling function $f(\Phi)$ of the scalar field to the Ricci scalar
are also expected to play a role in this analysis. Such an analysis will reveal whether
the scalar field may stabilise the black string over the bulk regime close to our
brane where it has a non-trivial profile. However, beyond the point where the
scalar-field energy-momentum tensor vanishes, we expect the Gregory-Laflamme
instability to set in, as in all other infinitely extended black-string 
solutions. The stability of the solutions for the
case of $M=0$ should also be carefully examined as the role of the singularity,
arising at the boundary of spacetime when $\Lambda \neq 0$, may be found
to be important. We hope to return to these questions soon.

\bigskip

{\bf Acknowledgements.} The research of T.N. was co-financed by Greece and the
European Union (European Social Fund- ESF) through the Operational Programme
``Human Resources Development, Education and Lifelong Learning'' in the context
of the project ``Strengthening Human Resources Research Potential via Doctorate
Research'' (MIS-5000432), implemented by the State Scholarships Foundation (IKY).
The research of N.P. was implemented under the
``Strengthening Post-Doctoral Research" scholarship program (grant no 2016-050-0503-7626)
by the Hellenic State Scholarship Foundation as part of the Operational Program
``Human Resources Development Program, Education and Lifelong Learning", co-financed
by the European Social Fund-ESF and the Greek government.

\appendix

\numberwithin{equation}{section}

\section{Curvature Invariant Quantities}
\label{App-Invar}

Employing the form of the five-dimensional line-element (\ref{metric}), we compute
the following scalar invariant quantities:
\eq$\label{4}
R=-8 A''-20 A'^2+\frac{2 e^{-2 A} \left(r \pa_r^2m+2 \pa_rm\right)}{r^2}\,,$
\begin{align}\label{5}
R_{MN}R^{MN}=2e^{-4 A} \left[e^{2 A} \left(A''+4 A'^2\right)-\frac{\pa_r^2m}{r}\right]^2&
+2\frac{e^{-4 A} \left[r^2 e^{2 A}\left(A''+4 A'^2\right)-2 \pa_rm\right]^2}{r^4}\nonum\\
&+16 \left(A''+A'^2\right)^2,
\end{align}
\begin{align}\label{6}
R_{MNKL}R^{MNKL}=&-\frac{8 e^{-2 A} A'^2 \left(r \pa_r^2m+2 \pa_rm\right)}{r^2}
+40 A'^4+16 A'' \left(A''+2 A'^2\right)\nonum\\[2mm]
&+4 e^{-4 A}\left[\frac{(\pa_r^2m)^2}{r^2}+\frac{4\left[2 (\pa_rm)^2+ 
\left(m-r \pa_rm\right) \pa_r^2m\right]}{r^4}+\frac{4 (3 m^2-4 r m \pa_rm)}{r^6}\right].
\end{align}


\section{A systematic methodology to express 
$\boldsymbol{\,_2F_1\left(\frac{3}{2}-\frac{\lam}{4},\frac{3}{2};\frac{5}{2};u^2\right)}$
in terms\\[2mm] of elementary functions when $\boldsymbol{\lam=4q}$, and 
$\boldsymbol{q \in} {\mathbb{Z}}^{>}$}
\label{hyper-analysis}
\vspace{1em}

Let us start with the simplest case of $\lam=4$, i.e. $q=1$. For simplicity, we will
use the variable $u^2=\frac{w-1}{w}$. Then, using the expansion of the hypergeometric
function given in Eq. (\ref{final-F}) and setting $\lam=4$, we readily obtain
\bal$
\label{hyper-lam-4}\,_2F_1\left(\frac{1}{2},\frac{3}{2};\frac{5}{2};u^2\right)=
\frac{3}{2\sqrt{\pi}}\sum_{n=0}^\infty \frac{\Gamma\left(n+\frac{1}{2}\right)}
{n+\frac{3}{2}}\frac{u^{2n}}{n!}=\frac{3}{2\sqrt{\pi}}\sum_{n=0}^\infty
\frac{2n-1}{2n+3}\,\Gamma\left(n-\frac{1}{2}\right)\frac{u^{2n}}{n!}$
where we have used the Gamma function property $\Gamma(1+z)=z\Gamma(z)$. In order to express
the above in terms of elementary functions, we observe the following
\bal$
\label{arcsin-sqrt}
\frac{\arcsin u}{u}-\sqrt{1-u^2}&=\,_2F_1\left(\frac{1}{2},\frac{1}{2};\frac{3}{2};u^2\right)
-\,_2F_1\left(-\frac{1}{2},1;1;u^2\right)\nonum\\[2mm]
&=\sum_{n=0}^\infty \left[\frac{\Gamma\left(n+\frac{1}{2}\right)}{(2n+1)\sqrt{\pi}}+
\frac{\Gamma\left(n-\frac{1}{2}\right)}{2\sqrt{\pi}}\right] \frac{u^{2n}}{n!}=
\sum_{n=1}^\infty \frac{2n}{2n+1}\,\frac{\Gamma\left(n-\frac{1}{2}\right)}{\sqrt{\pi}}\,
\frac{u^{2n}}{n!}\nonum\\[2mm]
&=\sum_{m=0}^\infty \frac{2(m+1)}{2m+3}\,\frac{\Gamma\left(m+\frac{1}{2}\right)}
{\sqrt{\pi}}\,\frac{u^{2(m+1)}}{(m+1)!}=\frac{1}{\sqrt{\pi}}\sum_{m=0}^\infty 
\frac{2m-1}{2m+3}\,\Gamma\left(m-\frac{1}{2}\right)\frac{u^{2(m+1)}}{m!}$
Note that, in the second sum of the second line of the above equation, we have changed
the lower value of $n$ from $n=0$ to $n=1$ since, due to the ($2n$) factor, this value
has a trivial contribution to the sum. Subsequently, we set $n=m+1$, and by rearranging
we arrived at the final result. Comparing now Eqs. \eqref{hyper-lam-4} and
\eqref{arcsin-sqrt}, we find that
\eq$ \label{B.3}\,_2F_1\left(\frac{1}{2},\frac{3}{2};\frac{5}{2};u^2\right)u^2=
\frac{3}{2}\left(\frac{\arcsin u}{u}-\sqrt{1-u^2}\right).$\vspace{1em}

Let us now address the more general case where $\lam=4q$, with $q=1+\ell$ where
$\ell$ a positive number. Then, applying again Eq. (\ref{final-F}), we may write
\bal$
\label{hyper-lam-4q}\,_2F_1\left(\frac{1}{2}-\ell,\frac{3}{2};\frac{5}{2};u^2\right)&=
\frac{3}{\sqrt{\pi}}\,\frac{(2\ell)!}{\ell!(-4)^\ell}\sum_{n=0}^\infty \frac{1}{2n+3}\,
\Gamma\left(n-\ell+\frac{1}{2}\right)\frac{u^{2n}}{n!}\,,$
where we have also used the property $\Gamma\left(-\ell+\frac{1}{2}\right)=\frac{\ell!(-4)^\ell\sqrt{\pi}}{(2\ell)!}$.
But, it also holds that
\bal$\label{gamma-1/2}
\Gamma\left(n-\frac{1}{2}\right)&=\frac{\Gamma\left(n-\ell+\frac{1}{2}+\ell\right)}{n-\frac{1}{2}}
=\frac{2}{2n-1}\left(n-\frac{1}{2}\right)\left(n-\frac{3}{2}\right)\cdots\left(n-\ell+\frac{1}{2}\right)\Gamma\left(n-\ell+\frac{1}{2}\right)\nonum\\[2mm]
&=2^{1-\ell}\,\frac{(2n-1)(2n-3)\cdots(2n-2\ell+1)}{2n-1}\,\Gamma\left(n-\ell+\frac{1}{2}\right).$
From Eqs. \eqref{arcsin-sqrt} and \eqref{gamma-1/2}, we then obtain
\gat$\label{arcsin-sqrt-ell}
\frac{\arcsin(u)}{u}-\sqrt{1-u^2}=\frac{2^{1-\ell}}{\sqrt{\pi}}\sum_{n=0}^\infty \frac{(2n-1)(2n-3)\cdots(2n-2\ell+1)}{2n+3}\Gamma\left(n-\ell+\frac{1}{2}\right)\frac{u^{2(n+1)}}{n!}.$
In what follows, we discuss also how the multiplication between even powers of $u$
and $\sqrt{1-u^2}$ can result in similar expansions as the one in Eq. \eqref{arcsin-sqrt-ell}.
The obtained expansions together with Eq. \eqref{arcsin-sqrt-ell} help us to express
the r.h.s. of Eq. \eqref{hyper-lam-4q} in terms of elementary functions. Thus, starting
from the relation
\eq$\sqrt{1-u^2}=-\frac{1}{2\sqrt{\pi}}\sum_{n=0}^\infty
\Gamma\left(n-\frac{1}{2}\right)\frac{u^{2n}}{n!}\,,$
we write, employing also Eq. (\ref{gamma-1/2}),
\bal$\label{z2-sqrt}
u^2\sqrt{1-u^2}&=-\frac{2^{1-\ell}}{2\sqrt{\pi}}\sum_{n=0}^\infty
\frac{(2n-1)(2n-3)\cdots(2n-2\ell+1)}{2n-1}\Gamma\left(n-\ell+\frac{1}{2}\right)
\frac{u^{2(n+1)}}{n!}.$
Similarly, we obtain
\bal$\label{z4-sqrt}
u^4\sqrt{1-u^2}&=-\frac{1}{2\sqrt{\pi}}\sum_{m=0}^\infty m\ \Gamma\left(m-\frac{3}{2}\right)
\frac{u^{2(m+1)}}{m!}=-\frac{1}{2\sqrt{\pi}}\sum_{m=0}^\infty \frac{2m}{2m-3}\ \Gamma\left(m-\frac{1}{2}\right)\frac{u^{2(m+1)}}{m!}\nonum\\[2mm]
&=-\frac{2^{2-\ell}}{2\sqrt{\pi}}\sum_{n=0}^\infty \frac{(2n-1)(2n-3)\cdots(2n-2\ell+1)\ n}{(2n-1)(2n-3)}\ \Gamma\left(n-\ell+\frac{1}{2}\right)\frac{u^{2(n+1)}}{n!}.$
Note that, in the first sum of the above expression, we set $m=n+1$ but retained the lower
value of the sum to be 0 due to the $m$ factor---we have also used, once again,
Eq. (\ref{gamma-1/2}). Continuing along the same lines, we obtain, for a general $\ell$,
the result 
\bal$\label{zell-sqrt}
u^{2\ell}\sqrt{1-u^2}&=
-\frac{1}{2\sqrt{\pi}}\sum_{m=\ell-1}^\infty m(m-1)\cdots(m-\ell+2)\ \Gamma\left(m-\ell+\frac{1}{2}\right)\frac{u^{2(m+1)}}{m!}\nonum\\[2mm]
&=-\frac{1}{2\sqrt{\pi}}\sum_{n=0}^\infty n(n-1)\cdots(n-\ell+2)\ \Gamma\left(n-\ell+\frac{1}{2}\right)\frac{u^{2(n+1)}}{n!}\,,$
where, now, we set $m=n+\ell-1$ and again reinstated the lower value of the sum to be 0
due to the multiplying factors that trivialise all terms with $n < \ell-1$. 

Comparing now the r.h.s's of Eqs. \eqref{arcsin-sqrt-ell}, \eqref{z2-sqrt} and
\eqref{zell-sqrt} with the r.h.s of Eq. \eqref{hyper-lam-4q},
we conclude that we may express the aforementioned hypergeometric function as
\bal$\label{hyper-lam-4q-final}
\,_2F_1\left(\frac{1}{2}-\ell,\frac{3}{2};\frac{5}{2};u^2\right)u^2=&
\alpha\left(\frac{\arcsin u}{u}-\sqrt{1-u^2} \right)\nonum\\[2mm]
&+\sqrt{1-u^2}\left(\beta_1\,u^2+\beta_2\,u^4+\dots+\beta_{\ell-1}\,u^{2(\ell-1)}+
\beta_\ell\,u^{2\ell}\right),$
where $\alpha, \beta_1, \dots, \beta_\ell$ are constant coefficients. These may be
determined by substituting the explicit expansions (\ref{hyper-lam-4q}),
\eqref{arcsin-sqrt-ell}, \eqref{z2-sqrt} and \eqref{zell-sqrt} on both sides of
the above equation and demanding its validity. Then, we obtain the following
relation for the coefficients $\alpha, \beta_1, \dots, \beta_\ell$, which must
be true for arbitrary $n\in\mathbb{Z}^{\geq}$,
\bal$
6\frac{(2\ell)!}{ \ell!\,(-4)^\ell}=&\hspace{0.5em}\alpha\,2^{2-\ell}\,(2n-1)(2n-3)\cdots(2n-2\ell+1)\nonum\\[1mm]
&-(2n+3)\bigg[\,\beta_1\,2^{1-\ell}\ \frac{(2n-1)(2n-3)\cdots(2n-2\ell+1)}{2n-1}\nonum\\[1mm]
&\hspace{5em}+\beta_2\,2^{2-\ell}\,\frac{(2n-1)(2n-3)\cdots(2n-2\ell+1)}{(2n-1)(2n-3)}
\,n\nonum\\[1mm]
&\hspace{5em}+\beta_3\,2^{3-\ell}\,\frac{(2n-1)(2n-3)\cdots(2n-2\ell+1)}{(2n-1)(2n-3)(2n-5)}\, n(n-1)\nonum\\[1mm]
&\hspace{5cm}\vdots\nonum\\[1mm]
&\hspace{5em}+\beta_{\ell-1}\,2^{-1}\,(2n-2\ell+1)\ n(n-1)\cdots(n-\ell+3)\nonum\\[1mm]
&\hspace{5em}+\beta_{\ell}\,\frac{(n+1)n(n-1)\cdots(n-\ell+2)}{n+1}\,\bigg]
\label{alpha-beta}$
The above equation leads to a system of $\ell+1$ linear equations with $\ell+1$ variables
from which the unknown coefficients $\alpha, \beta_1, \dots, \beta_\ell$ may easily be
derived.

\bibliographystyle{unsrt}
\bibliography{Bibliography}
\addcontentsline{toc}{chapter}{\numberline{}References}

\end{document}